\documentclass[structabstract]{aa}
\usepackage{txfonts}
\usepackage{aas_macros}
\usepackage{natbib}
\bibpunct{(}{)}{;}{a}{}{,} 
\usepackage{version}
\begin{document}
\setlength{\topmargin}{-1cm}
\title{Variations on a theme -- the evolution of hydrocarbon solids:}
\subtitle{I. Compositional and spectral modelling -- the eRCN and DG models \\[0.2cm] 
{\small Published: A\&A, 540, A1 (2012) -- DOI: 10.1051/0004-6361/201117623 \\[0.2cm] 
Data files are available at the CDS via \\[-0.1cm] 
http://cdsarc.u-strasbg.fr/viz-bin/qcat?J/A+A/545/C2 and http://cdsarc.u-strasbg.fr/viz-bin/qcat?J/A+A/545/C3}}

\author{A.P. Jones \inst{1,2}}
           
    \institute{Institut d'Astrophysique Spatiale, CNRS/Universit\'e Paris Sud, UMR\,8617, Universit\'e Paris-Saclay, Orsay F-91405, France\\
               \email{ Anthony.Jones@ias.u-psud.fr}              }

    \date{Received 4 July 2011 / Accepted 30 October 2011}

   \abstract
{The compositional properties of hydrogenated amorphous carbons are known to evolve in response to the local conditions.}
{To present a model for low-temperature, amorphous hydrocarbon solids, based on the microphysical properties of random and defected networks of carbon and hydrogen atoms, that can be used to study and predict the evolution of their properties in the interstellar medium.}
{We adopt an adaptable and prescriptive approach to model these materials, which is based on a random covalent network (RCN) model, extended here to a full compositional derivation (the eRCN model), and a defective graphite (DG) model for the hydrogen poorer materials where the eRCN model is no longer valid.}
{We provide simple expressions that enable the determination of the structural, infrared and spectral properties of amorphous hydrocarbon grains as a function of the hydrogen atomic fraction, $X_{\rm H}$. Structural annealing, 
resulting from hydrogen atom loss, results in a transition from H-rich, aliphatic-rich to H-poor, aromatic-rich materials.}
{The model predicts changes in the  optical properties of hydrogenated amorphous carbon dust in response to the likely UV photon-driven and/or thermal annealing processes resulting, principally, from the radiation field in the environment. 
We show how this dust component will evolve, compositionally and structurally in the interstellar medium in response to the local conditions. }

   \keywords{Interstellar Medium: dust, emission, extinction -- 
          Interstellar Medium: general}

   \maketitle

\section{Introduction}

In this paper we will refer to an amorphous hydrocarbon particle as any finite-sized,  macroscopically-structured ({\it {\it i.e.}} a contiguous network of atoms),  solid-state material consisting solely of carbon and hydrogen atoms.  There is now a wealth of laboratory data on such materials \citep[{\it e.g.},][]{1984JAP....55..764S,1986AdPhy..35..317R,1988Sci...241..913A,1987PhRvB..35.2946R,1995ApJS..100..149M,1997ApJ...490L.175S,2000ApJ...543L..85G,2004PhilTransRSocLondA..362.2477F,2005ApJ...620L.135D,2008ApJ...672L..81H} and also  a number of theoretical models which interpret their inherent structures \citep[{\it e.g.},][]{1986AdPhy..35..317R,1987PhRvB..35.2946R,1988JVST....6.1778A,1988PMagL..57..143R,1990JAP....67.1007T,1990MNRAS.247..305J,1995ApJ...445..240D,1997ApJ...476..184D,2005ApJ...626..933D} and it is upon these data and models that we shall rely in constructing  a model for the evolution of hydrocarbon grains in the interstellar medium (ISM).

It has long been known that laboratory hydrocarbon solids darken upon exposure to ultraviolet (UV) light and to heating \citep[{\it e.g.},][]{1985OpEffinAS.120..258I,1984JAP....55..764S}, and that this photo-darkening of the materials is accompanied by a decrease in the band gap or optical gap energy, $E_{\rm g}$. It is this property that is at the heart of the inherent variability of the  optical properties of such solid materials, and that is of prime interest in the context of the applicability of these properties in unravelling  the histories of hydrocarbon grains in the ISM \cite[{\it e.g.},][]{1996MNRAS.283..343D}.

The evolution of hydrocarbon materials is therefore of key interest for the study of interstellar dust properties because it appears that graphite is no longer a tenable candidate to explain the dust properties associated with carbon-rich interstellar dust \citep[{\it e.g.},][]{2004A&A...423L..33D,2005A&A...432..895D,2008A&A...492..127S,2011A&A...525A.103C}. Amorphous hydrocarbons, however, do appear to be a widespread component of interstellar dust in galaxies \citep[{\it e.g.},][]{2004A&A...423..549D}. These materials are consistent with the observed infrared absorption bands \citep[{\it e.g.},][]{2004A&A...423L..33D,2007A&A...476.1235D} and show temperature-dependent luminescence  \citep[{\it e.g.},][]{1986AdPhy..35..317R} consistent with the observed interstellar luminescence in the red, the so-called extended red emission \citep[ERE,][]{1997ApJ...482..866D,2001ApJ...553..575D,2005A&A...432..895D,2010A&A...519A..39G} and with the likely irradiation effects and the associated aliphatic to aromatic transformation in the ISM \citep[{\it e.g.},][]{1990MNRAS.247..305J,2004A&A...423..549D,2004A&A...423L..33D,2008A&A...490..665P,2008ApJ...682L.101M,2011A&A...529A.146G}, in circumstellar regions \citep[{\it e.g.},][]{2003ApJ...589..419G,2007ApJ...662..389G,2007ApJ...664.1144S,2008A&A...490..665P}, in interplanetary dust particles \citep[IDPs,][]{2006A&A...459..147M} and solar system organics \citep{2011A&A...533A..98D}. 

The work presented here was conceived with the application of the Random Covalent Network (RCN) theory for the structure of amorphous hydrocarbon materials \citep{1979PhRvL..42.1151P,1980JNS...42...87D,1983JNCS...57..355T,1988JVST....6.1778A,1990MNRAS.247..305J} to the astrophysical context \citep{1990MNRAS.247..305J} and this series of follow-up papers has therefore been some 21 years in gestation. 

In this paper we study the compositional, structural and spectral properties of amorphous hydrocarbon materials and apply this to the nature of low-temperature particles in the interstellar medium.

The paper is organised as follows:  in sections \S\,\ref{sect_struct_models} and \ref{sect_eRCN_DG_spectra} we present the eRCN and defective graphite (DG) models for the structural evolution of hydrogenated amorphous carbons and their associated spectra, in \S\,\ref{sect_astro_implications} we discuss the astrophysical implications for the evolution of carbonaceous dust in the ISM,  in \S\,\ref{sect_predictions} we summarise the model predictions, in \S\,\ref{sect_limitations} we discuss the limitations of the eRCN/DG models, and in \S\,\ref{sect_conclusions} we present out conclusions. 

In the ensuing, associated papers we study the compositional and wavelength-dependent optical properties  of amorphous hydrocarbons within the astrophysical context (paper~II) and the r\^ole of size effects on those optical properties.

\section{Structural models for amorphous hydrocarbons}
\label{sect_struct_models}

Hydrocarbon materials span the entire compositional range between the ordered diamond-like solids and graphite-like solids (CH$_{\rm n}$, where $n \approx 0$) and organic polymers (CH$_{\rm n}$, where $n \approx 2$), but it is  the amorphous hydrocarbon materials that exist between these various forms that are of most interest here (CH$_{\rm n}$, where $0 \leq n \leq 2$). This is because amorphous hydrocarbons exhibit properties that can change in response to their irradiation and  thermal histories \citep[{\it e.g.},][]{1990MNRAS.247..305J}. 

The crucial parameters that define the structures ({\it {\it i.e.}} the short-range order) of amorphous hydrocarbon materials are the carbon atom bonding, in particular the ratio of the $sp^3$ and $sp^2$ fractions, and the hydrogen content \citep{1986AdPhy..35..317R,1987PhRvB..35.2946R,1988JVST....6.1778A,1990MNRAS.247..305J}. Of particular relevance is the way in which the $sp^3$ and $sp^2$ carbon atoms cluster, as indeed they must if they are to preserve the required atomic configurations \citep{1988JVST....6.1778A,1990MNRAS.247..305J}. Hydrocarbon solids therefore consist of domains of diamond-like or aliphatic ($sp^3$ bonded) and graphite-like or aromatic ($sp^2$ bonded) carbon intimately connected in networks that lack any long-range order but that are bridged by aliphatic and olefinic structures. The domains may consist of only a few to a few tens of atoms and contain both C and H atoms. At the short-range or atomic level there is some regularity in the networks due to the geometrical constraints imposed by the bonding of the carbon atoms as a function of their  hybridisation state {\it {\it i.e.}} $sp^3$ or $sp^2$, over the long-range there is no significant order. 

In order to understand the structures of hydrocarbon materials a particularly useful set of models have been the random covalent network (RCN) theories \citep{1979PhRvL..42.1151P,1980JNS...42...87D,1983JNCS...57..355T,1988JVST....6.1778A,1990MNRAS.247..305J}. These use the average nearest-neighbour bonding environment to constrain a macroscopic RCN of carbon and hydrogen atoms. The network is completely constrained when the number of constraints per atom (related to the coordination number) is equal to the number of mechanical degrees of freedom \citep[the dimensionality,][]{1979PhRvL..42.1151P}. Within a network the formation of bonds leads to increased stability, however, in a random network directed bonds lead to strain energy due to distortions in the bond lengths and angles. The optimal network is one that just balances these two effects. 

In the following sections we summarise the salient points of the RCN theories and extend the model to more complex compositions.

\subsection{The basic Random Covalent Network (RCN) model}
\label{RCN_basics}

In order to aid the reader, we here summarise the essential elements of the RCN formalism, which is described in detail elsewhere  \citep{1979PhRvL..42.1151P,1980JNS...42...87D,1983JNCS...57..355T,1988JVST....6.1778A,1990MNRAS.247..305J}.  

If in a network $m$ is the coordination number of a given atom (1 for hydrogen, and 2, 3 and 4 for $sp^1$, $sp^2$ and $sp^3$ carbon atoms,  respectively) then the number of constraints for a given atomic site, $N_{\rm con}$, in a three-dimensional network, where the bonding is dominated by nearest  neighbour, directed valence bonds is, as given by \cite{1980JNS...42...87D}, 
\[
N_{\rm con} = \frac{1}{2} m^2 \ \ \ \ \ \ \ \ \ \  {\rm if}\ \ m \leq 2,
\]
\begin{equation}
N_{\rm con} = \frac{5}{2} m-3 \ \ \ \ \    {\rm if}\ \ m \geq 2,
\label{eq_Nconi}
\end{equation}
and for a completely constrained network consisting of several types of atom of coordination number $i$ and atomic fraction $x_i$,
\begin{equation}
\sum_i x_i \, N_{{\rm con},i} = 3.
\label{xiNcon_sum_def}
\end{equation}
Hydrocarbon materials can be considered to consist primarily of only  hydrogen atoms, and $sp^2$ and $sp^3$ carbon atoms \citep{1986AdPhy..35..317R,1990MNRAS.247..305J}. Any $sp^1$ hybridised carbon atoms are of very low abundance in these materials and are, for simplicity, ignored in the RCN model. In this case two equations then relate the atom fractions for each component, $x_1$ (H), $x_3$ ($sp^2$) and $x_4$ ($sp^3$),  respectively, {\it i.e.},
\begin{equation}
x_1 + x_3 + x_4 = 1
\label{atom_frac_sum}
\end{equation}
\begin{equation}
0.5x_1 + 4.5x_3 + 7x_4 = 3,
\label{xiNcon_sum}
\end{equation}
and with some simple algebra we then have that 
\begin{equation}
x_3 = \frac{(8-13X_{\rm H})}{5} \equiv X_{sp^2} \ \ \ \ {\rm and} 
\end{equation}
\begin{equation}
x_4 = \frac{(8X_{\rm H}-3)}{5} \ \ \ \equiv X_{sp^3} 
 \end{equation}
where $X_{\rm H}$ is the atomic fraction of hydrogen ($\equiv x_1$). 
The ratio, $R$, of the $sp^3$ and $sp^2$ atom fractions and the average coordination number, $\bar{m}$, is then, following \cite{1988JVST....6.1778A},
\begin{equation}
R \equiv \frac{X_{sp^3}}{X_{sp^2}} = \frac{(8X_{\rm H}-3)}{(8-13X_{\rm H})},  
\label{RvsXH_simple}
\end{equation}
\begin{equation}
\bar{m} = X_{\rm H} + 3X_{sp^2} + 4X_{sp^3} = \frac{(12-2X_{\rm H})}{5} = 2.40 - 0.40 \, X_{\rm H},
\label{ave_coord_no_simple}
\end{equation}
where we now replace the fundamental quantities $x_3$ and $x_4$, the  $sp^2$ and $sp^3$  carbon atom fractions, respectively, by the more explicit symbols $X_{sp^2}$ and $X_{sp^3}$. 
The average carbon atom coordination number is obtained by considering only the $X_{sp^2}$ and $X_{sp^3}$ fractions in Eq.~(\ref{ave_coord_no_simple}) and then re-normalising to the carbon atom fraction $(1-X_{\rm H})$, {\it i.e.},
\begin{equation}
\bar{m_{\rm C}} = \frac{3X_{sp^2} + 4X_{sp^3}}{(1-X_{\rm H})} = \frac{(12-7X_{\rm H})}{5(1-X_{\rm H})}  = \frac{2.40 - 1.40 \, X_{\rm H}}{(1-X_{\rm H})} . 
\label{C_coord_no_simple}
\end{equation}
From Eq.~(\ref{RvsXH_simple}) we can see that for $sp^3$ bonding only, {\it i.e.} setting $X_{sp^2} = 0$, we have $X_{\rm H} = 8/13 = 0.615$. For hydrogen concentrations greater than 0.615 the number of bonds is insufficient to use up all the degrees of freedom and the network will be `under-constrained'. For $sp^2$ bonding only we can similarly set $X_{sp^3} = 0$ and in this case we have $X_{\rm H} = 3/8 = 0.375$. For hydrogen concentrations less than 0.375 the network would be `over-constrained'. In general, if the number of constraints per atom is greater than the number of dimensions then the network will re-construct in some way and small clusters violating this simple RCN model would be expected to appear. From the limits of the validity of Eq.~(\ref{RvsXH_simple}) we see that these RCNs can therefore only exist over the hydrogen content range $0.38 \lesssim X_{\rm H} \lesssim 0.62$. 

\begin{figure}
 \resizebox{\hsize}{!}{\includegraphics{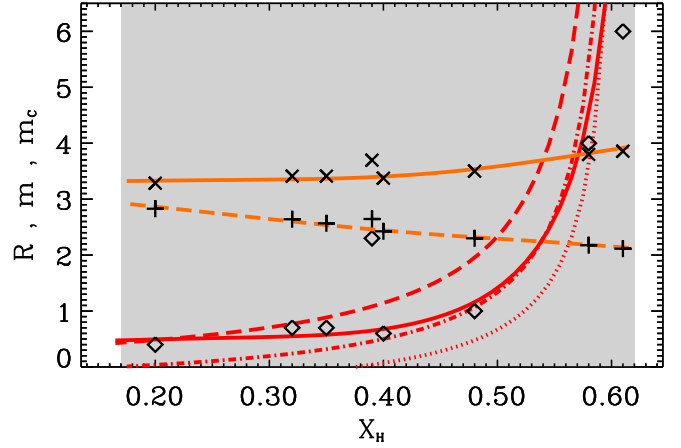}}
 \caption{Carbon atom ratio $R$ as a function of $X_{\rm H}$.  Dotted line: basic RCN model (Eq.~\ref{RvsXH_simple}), short-dashed-dotted line: C$_2$ $sp^2$ clusters (Eq.~\ref{RvsXH_fit1}), long-dashed line: all $sp^2$ carbon atoms in C$_6$ aromatic clusters (Eq.~\ref{RvsXH_fit2_allsp2C6}) and solid line: multi-C$_6$ ring aromatic clusters (Eq.~\ref{RvsXH_fit3}) and methyl groups (Eq.~\ref{RvsXH_fit_CH3}). The upper and flatter solid  lines show the mean atomic coordination number, $\bar{m}$ (dashed), and the mean carbon atom coordination number, $\bar{m}_C$ (solid) for the full model. Diamonds and plus signs [crosses]; laboratory data points for $R(X_{\rm H})$ and $\bar{m}(X_{\rm H})$ [$\bar{m}_C(X_{\rm H})$], respectively, taken from \cite{1988JVST....6.1778A} and references therein. }
\label{RvsXH}
\end{figure}

Strictly, the above description is only valid for systems where the triply coordinated $sp^2$ carbon atoms are randomly distributed. However, $sp^2$ sites must pair up into, at least, ethylenic units, {\it i.e.}, $>$C=C$<$, with a coordination number of 4 \citep{1988JVST....6.1778A} and these structures are observed in the IR spectra of a-C:H \citep{1983SolidStComm..48..105D,1983ApPhL..42..636D}. The equations governing the RCN network with paired $sp^2$ sites, analogous to Eqs.~(\ref{atom_frac_sum}) to (\ref{C_coord_no_simple}), are then,
\begin{equation}
X_{\rm H} + X_{sp^2} + X_{sp^3} = 1,
\label{atom_frac_sum_C2}
\end{equation}
\begin{equation}
0.5X_{\rm H} + 7 ( 0.5 X_{sp^2}) + 7X_{sp^3} = 3,
\label{xi_Ncon_sum_C2}
\end{equation}
\begin{equation}
X_{sp^3} = \frac{(6X_{\rm H}-1)}{7},
\end{equation}
\begin{equation}
X_{sp^2} = \frac{(8-13X_{\rm H})}{7}, 
\end{equation}
\begin{equation}
R = \frac{(6X_{\rm H}-1)}{(8-13X_{\rm H})},
\label{RvsXH_fit1}
\end{equation}
\begin{equation}
\bar{m} = X_{\rm H} + 3X_{sp^2} + 4X_{sp^3} = \frac{(20-8X_{\rm H})}{7}  \ \ = 2.86 - 1.14 \, X_{\rm H}, 
\label{ave_coord_no_C2}
\end{equation}
\begin{equation}
\bar{m_{\rm C}} = \frac{3X_{sp^2} + 4X_{sp^3}}{(1-X_{\rm H})}  \ \ \ \ \ \  = \frac{(20-15X_{\rm H})}{7(1-X_{\rm H})} = \frac{2.86 - 2.14\, X_{\rm H}}{(1-X_{\rm H})}, 
\label{ave_coord_no_C2_C}
\end{equation}
and in this case the above expressions are valid for $0.17 \lesssim X_{\rm H} \lesssim 0.62$. This range, derived using a rather simple model, is in excellent agreement with laboratory data \citep{1988JVST....6.1778A} and indicates the compositions for which completely constrained hydrocarbon networks can exist. In Fig.~\ref{RvsXH} we show $R$ as a function of $X_H$ for the models presented in this section and calculated using Eqs.~(\ref{RvsXH_simple}, dotted line) and (\ref{RvsXH_fit1}, short dash-dotted line). 

For RCN networks consisting of H atoms, $sp^3$ carbon atoms and C$_2$ $sp^2$ carbon ethylenic ($>$C=C$<$) groups, and where $0.17 \lesssim X_{\rm H} \lesssim 0.62$, the range of values for the mean atomic coordination number, $\bar{m}$, given by Eq.~(\ref{ave_coord_no_C2}), is $2.67 \lesssim \bar{m} \lesssim 2.15$. For an H-poor, aromatic-rich structure the limiting lower limit must be close to three, while the upper limit should approach that for a (CH$_2$)$_n$ polymer, {\it i.e.}, $\bar{m} = \frac{1}{3} \times 4 + \frac{2}{3} \times 1 = 2$, which is clearly consistent with the H-rich limit of 2.15 derived here. The mean carbon atom coordination, $\bar{m_{\rm C}}$, given by Eq.~(\ref{ave_coord_no_C2_C}), is equivalently $3.00 \lesssim \bar{m_{\rm C}} \lesssim 4.03$, which reproduces the expected values, {\it i.e.}, 3.0 for $sp^2$-only and 4.0 for $sp^3$-only structures.

\cite{1995ApJ...445..240D} extended the RCN model to include $sp^3$ C$_6$H$_{10}$ `diamond-like' clusters with four aliphatic CH$_2$ groups and two tertiary CH groups. We note, however, that the smallest aliphatic cluster of this type is actually the adamantane-like cluster C$_{10}$H$_{16}$, which actually contains six aliphatic CH$_2$ groups and four tertiary CH groups.

\subsection{The extended RCN model (eRCN)}
\label{sect_extended_RCN}

With the aim of simplifying our further, and following, extensions to the basic RCN model(s), and in order to facilitate the algebraic manipulations, we now parameterise Eqs.~(\ref{xiNcon_sum}) and (\ref{xi_Ncon_sum_C2}) to: 
\begin{equation}
aX_{\rm H} + bX_{sp^2} + cX_{sp^3} = 3 
\end{equation}
and then, following some simple algebra, derive generalised expressions for the $sp^3$ and $sp^2$ carbon atom fractions ($X_{sp^3}$ and $X_{sp^2}$), their ratio $R \equiv X_{sp^2}/X_{sp^3}$,  and the average atomic coordination number, $\bar{m}$, {\it i.e.}, 
\begin{equation}
X_{sp^3} = \frac{(8X_{\rm H}-3)}{5} = \frac{(b-a)X_{\rm H}-(b-3)}{(c-b)} = \frac{R(1-X_{\rm H})}{(1+R)},
\label{gen_Xsp3_abc}
\end{equation}
\begin{equation}
X_{sp^2} = \frac{(8-13X_{\rm H})}{5} = \frac{(c-3)-(c-a)X_{\rm H}}{(c-b)} = \frac{(1-X_{\rm H})}{(1+R)},
\label{gen_Xsp2_abc} 
\end{equation}
\begin{equation}
R  = \frac{(b-a)X_{\rm H}-(b-3)}{(c-3)-(c-a)X_{\rm H}}, 
\label{gen_R_abc}
\end{equation}
\[
\bar{m} = X_{\rm H} + 3X_{sp^2} + 4X_{sp^3}, 
\]
\[
\ \ \ \  = X_{\rm H} + 3\left[ \frac{(c-3)-(c-a)X_{\rm H}}{(c-b)} \right] + 4\left[ \frac{(b-a)X_{\rm H}-(b-3)}{(c-b)} \right], 
\]
\[
\ \ \ \   = X_{\rm H} + \frac{3(1-X_{\rm H})}{(1+R)} + \frac{4R(1-X_{\rm H})}{(1+R)}, 
\]
\begin{equation}
\ \ \ \  = \frac{(3-2X_{\rm H})+(4-3X_{\rm H})R}{(1+R)}.
\label{gen_m_abc}
\end{equation}
In the above expressions, and in many of those that follow, the factor $(1-X_{\rm H})$ often appears; this is simply the total carbon atomic fraction,  {\it i.e.}, $(1-X_{\rm H}) \equiv (X_{sp^2} + X_{sp^3}) \equiv X_{\rm C}$. However, given that {\em the} key parameter for RCN, and the following eRCN, models is $X_{\rm H}$, the substitution of $X_{\rm C}$ for $(1-X_{\rm H})$ is not particularly useful or informative and we chose not to make it here.

It is clear that hydrogen atoms do not contribute to the connectivity of a network and, if we neglect their contribution and re-normalise to the total number of carbon atoms, we derive the general formula for the mean carbon atom coordination number, $\bar{m}_C$, for the carbon framework, {\it {\it i.e.}},
\[
\bar{m}_{\rm C} = 3\left(\frac{X_{sp^2}}{1-X_{\rm H}}\right) + 4\left(\frac{X_{sp^3}}{1-X_{\rm H}}\right) = \frac{(3+4R)}{(1+R)}, 
\]
\begin{equation}
\ \ \ \ \ \  = 3 \left[ \frac{(c-3)-(c-a)X_{\rm H}}{(c-b)(1-X_{\rm H})} \right] + 4 \left[ \frac{(b-a)X_{\rm H}-(b-3)}{(c-b)(1-X_{\rm H})} \right]. 
\label{gen_mC_abc}
\end{equation}

In most RCN systems we find that $a=\frac{1}{2}$ and $c=7$  and we can therefore further simplify the above expressions to, 
\begin{equation}
X_{sp^3} = \frac{(2b-1)X_{\rm H}-(2b-6)}{2(7-b)},  
\end{equation}
\begin{equation}
X_{sp^2} = \frac{(8-13X_{\rm H})}{2(7-b)}, 
\end{equation}
\begin{equation}
R  = \frac{(2b-1)X_{\rm H}-(2b-6)}{(8-13 X_{\rm H})}, 
\label{gen_R_b}
\end{equation}
\[
\bar{m} = \frac{2(7-b)X_{\rm H} + 3[8-13X_{\rm H}] + 4[(2b-1)X_{\rm H}-(2b-6)]}{2(7-b)}, 
\]
\begin{equation}
\ \ \ \  = \frac{(48-8b)-(29-6b)X_{\rm H}}{2(7-b)}, 
\label{gen_m_b}
\end{equation}
\[
\bar{m}_{\rm C}   = \frac{3(8-13X_{\rm H}) + 4[(2b-1)X_{\rm H}-(2b-6)]}{2(7-b)}, 
\]
\begin{equation}
\ \ \ \ \ \  = \frac{(48-8b)-(43-8b)X_{\rm H}}{2(7-b)(1-X_{\rm H})}. 
\label{gen_mC_b}
\end{equation}

In our calculations we always use the appropriate expressions for $a$, $b$ and $c$, substituted into 
Eqs.~(\ref{gen_Xsp3_abc}) to (\ref{gen_mC_abc}), so as to avoid any algebraic errors that may have crept into the derivation of the more explicit equations presented in the following sections.

\subsubsection{Aromatic structures in eRCNs}
\label{section_aromatics}

The RCN formalism was further extended to the more realistic case where the $sp^2$ carbon atoms are allowed to cluster into C$_6$ aromatic ring units  \cite{1990MNRAS.247..305J}, which is consistent with the fact that even-numbered C atom clusters are more stable and therefore favoured \citep{1986AdPhy..35..317R}. This introduces six-fold coordination sites, $X_6$, with $N_{\rm con} = 12$. The $X_6$ sites exhibit only short-range order and their inclusion in the RCN scheme is therefore valid. The equations for atomic conservation and the summed atomic constraints then become
\begin{equation}
X_{\rm H} + X_{sp^2} + X_{sp^3} + X_6 = 1, 
\end{equation}
\begin{equation}
0.5X_{\rm H} + 7 (0.5 X_{sp^2}) + 7X_{sp^3} + 12(0.167X_6) = 3.
\end{equation}
If $f$ is the fraction of the $sp^2$ carbon atoms in the C$_6$ units then $X_6 \equiv f X_{sp^2}$ and the two above equations become
\begin{equation}
X_{\rm H} + (1-f)X_{sp^2} + X_{sp^3} + fX_{sp^2} = X_{\rm H} + X_{sp^2} + X_{sp^3}  = 1, 
\end{equation}
\[
0.5X_{\rm H} + 7 [0.5 (1-f) X_{sp^2}] + 12(0.167 f X_{sp^2}) + 7X_{sp^3}  = 3, 
\]
\begin{equation}
0.5X_{\rm H} + 7 [0.5 (1-f) + 2 f ] X_{sp^2} + 7X_{sp^3}  = 3.
\end{equation}
The equations for $R$ and $\bar{m}$ are then, for $a = 0.5$, $b = 7 [0.5 (1-f) + 2 f ] = 0.5 (7- 3f)$ an $c = 7$, 
\begin{equation}
R = \frac{(6X_{\rm H}-1)+3f(1-X_{\rm H})}{(8-13X_{\rm H})}, 
\label{RvsXH_fit2}
\end{equation}
\begin{equation}
\bar{m} = \frac{(20-8X_{\rm H})+3f(4-3X_{\rm H})}{(7+3f)}, 
\end{equation}
\begin{equation}
\bar{m_{\rm C}} = \frac{(20-15X_{\rm H})+12f(1-X_{\rm H})}{(7+3f)(1-X_{\rm H})}. 
\end{equation}
These equations reduce to those given previously when $f=0$. The case for $f=1$, {\it i.e.} all the $sp^2$ carbon atoms in C$_6$ rings, gives
\begin{equation}
R = \frac{(3X_{\rm H}+2)}{(8-13X_{\rm H})}, 
\label{RvsXH_fit2_allsp2C6}
\end{equation}
\begin{equation}
\bar{m} = \frac{(32 -17X_{\rm H})}{10}, 
\end{equation}
\begin{equation}
\bar{m_{\rm C}} = \frac{(32 - 27X_{\rm H})}{10(1-X_{\rm H})}, 
\end{equation}
which leads to a good fit to the laboratory data for amorphous hydrocarbons for low values of $X_{\rm H}$ (see Fig.~\ref{RvsXH}). However, at higher hydrogen contents the fit is less good because too much order is introduced into the network due to the `forced' insertion of unrealistic aromatic ring structures into a predominantly aliphatic network.

The RCN model can be further adapted to allow for the presence of multi-C$_6$ ring aromatic domains of $sp^2$ carbon atoms, {\it i.e.}, the introduction of polycyclic aromatics \citep{1990MNRAS.247..305J}. This formalism can be summarised in the following set of equations for the atomic conservation, the summed atomic constraints, the ratio $R$ and the average atomic coordination numbers, respectively,
\begin{equation}
X_{\rm H} + X_{sp^2} + X_{sp^3} = 1,
\label{RvsXH_fit3_frac}
\end{equation}
\begin{equation}
0.5X_{\rm H} + \{ 7 [0.5(1-f)] + Zf \} X_{sp^2} + 7X_{sp^3} = 3.
\end{equation}
In the interests of space-saving in the following, we define an aromatic carbon component parameter $Y_f = (7-2Z)f$. 
Here we now have $a = 0.5$, $b = (7 [0.5(1-f)] + Zf) = 0.5[7-(7-2Z)f] = 0.5(7-Y_f)$ and $c = 7$ and then
\begin{equation}
R = \frac{(6X_{\rm H}-1)-Y_f(1-X_{\rm H})}{(8-13X_{\rm H})},
\label{RvsXH_fit3}
\end{equation}
\begin{equation}
\bar{m} = \frac{(20-8X_{\rm H})+Y_f(4-3X_{\rm H})}{(7+Y_f)}, 
\label{RvsXH_fit3_meanm}
\end{equation}
\begin{equation}
\bar{m_{\rm C}} = \frac{(20-15X_{\rm H})+4Y_f(1-X_{\rm H})}{(7+Y_f)(1-X_{\rm H})}.
\label{RvsXH_fit3_meanmC}
\end{equation}
Here $Z$ is a function of the number of six-fold rings, $N_{\rm R}$, in the polycyclic aromatic clusters and is the number of constraints per carbon atom for the given cluster, {\it i.e.},
\begin{equation}
Z = \frac{N_{{\rm con},N_{\rm R}}} {n_{\rm C}},  
\end{equation}
where $n_{\rm C}$ is the number of carbon atoms per aromatic cluster. The relevant expressions for $Z$ for linear and compact aromatic clusters are given in Appendix~\ref{aromatics}. As above, $f$ is the fraction of $sp^2$ carbon atoms in the aromatic clusters or domains and is an empirically-derived function of $X_{\rm H}$. 
Here we have adopted an expression, which differs from that of \cite{1990MNRAS.247..305J}, in that we use a smooth transition from the maximum value of $f$ ($f_{\rm max}$) at low $X_{\rm H}$ to $f \approx 0$ at high $X_{\rm H}$ using 
\begin{equation}
f  =  f_{\rm max} / \{ {\rm e}^{(X_{\rm H}-X_{\rm Hc})/\delta} + 1 \}, 
\end{equation}
where we take $f_{\rm max}$ = 0.6 as the maximum fraction of $sp^2$ carbon atoms in the aromatic clusters, $X_{\rm Hc} = 0.33$  as a critical hydrogen atom fraction and $\delta = 0.07$, which parameterises the steepness of the transition between high and low $f$. It appears that our results are not particularly sensitive to the exact form of $f$, provided that the shape of the transition is physically `reasonable'.

If the simplifying assumption is again made that $f=1$ then it is an easy matter to calculate $R$ as a function of X$_{\rm H}$ for a wide range of aromatic clusters (PAH-like species). Rather surprisingly when this is done the results do not differ significantly from the single six-fold ring case, even for 4--5 ring linear and compact aromatic clusters \citep{1990MNRAS.247..305J}. Therefore it seems that limiting the consideration to the case for only benzene-like rings in hydrocarbon RCNs would be a reasonable assumption to make. This is supported by the observation that the $\pi-\pi^*$ transition in amorphous hydrocarbons occurs at 6.5~eV which is close to the prominent $^1A_{1\rm g}-^1E_{1\rm u}$ transition of benzene at 7.0~eV \citep{1984PhRvB..30.4713F}, therefore implying that a common aromatic cluster in amorphous hydrocarbons is the six-fold ring.

This RCN model extended to include polycyclic aromatic clusters (eRCN) has been shown to give a good fit to experimental data in the range $0.2 < X_{\rm H} < 0.6$ \citep{1990MNRAS.247..305J}, the range over which reliable experimental data are available \citep[][see Fig.~\ref{RvsXH}]{1988Sci...241..913A}. However, at values of $X_{\rm H} < 0.2$ the RCN approach is not valid because these hydrocarbons probably contain significant long-range order (aromatic carbon domains) and are therefore no longer true random networks. 

In Fig.~\ref{RvsXH} we show $R$ as a function of $X_{\rm H}$ as given in Eqs.~(\ref{RvsXH_simple}), (\ref{RvsXH_fit1}), (\ref{RvsXH_fit2_allsp2C6}) and (\ref{RvsXH_fit3}) for the basic RCN model and its more sophisticated eRCN derivatives. 

It is interesting to compare the atomic coordination numbers for the best-fit model and Fig.~\ref{RvsXH} therefore shows the mean carbon-plus-hydrogen and carbon-atom-only coordination numbers, $\bar{m}$ and $\bar{m_{\rm C}}$, respectively, derived from our most sophisticated RCN-derived model that includes aromatic clusters, and methyl groups (see the following section).

\subsubsection{The inclusion of methyl groups (--{\rm CH}$_3$) into eRCNs}
\label{inc_methyl}

Some years ago it was shown that modelling the infrared spectra of a-C:H in the C--H stretching and bending region at $2.5-10~\mu$m ($1000-4000$ cm$^{-1}$) required the, previously un-reported, presence of a significant methly group, --CH$_3$, concentration \citep{1998JAP....84.3836R}. This kind of grouping was not considered in the original RCN models \citep{1980JNS...42...87D,1988JVST....6.1778A,1990MNRAS.247..305J}. The presence of methy groups therefore requires a fundamental re-thinking and `re-tuning', of the current RCN/eRCN models. This primarily arises because the methyl groups, like hydrogen atoms, are structure-terminating components with unit coordination number. There is therefore a degeneracy between H atoms and --CH$_3$ groups in a RCN network.

In order to take into account the presence of methyl groups in the network we consider that a certain fraction, $\phi$, of the $sp^3$ carbon atom sites are replaced by --CH$_3$ groups, effectively inserting an extra, `non-linking' --CH$_2$-- group into the some of the C--H bonds in the a-C:H structure. In this case the number of $sp^3$ carbon atoms in the structure, $X_{sp^3}$, is not affected but there is resultant increase in the hydrogen atom fraction by a factor of $\phi X_{sp^3}$. This corresponds to the addition of one extra hydrogen atom per $sp^3$ methyl carbon atom, rather than three additional hydrogen atoms as might be expected. This is because in the eRCN model there are already one or two hydrogen atoms attached to every $sp^3$ C atom for  $X_{\rm H} \geq 0.5$. Thus, the eRCN equations, including the presence of $sp^2$ ethylenic C$_2$ clusters (Eqs.~\ref{atom_frac_sum_C2}, \ref{xi_Ncon_sum_C2}, \ref{RvsXH_fit1}, \ref{ave_coord_no_C2} and \ref{ave_coord_no_C2_C}), can be re-written in the form
\begin{equation}
X_{\rm H} + X_{sp^2} + X_{sp^3} = 1,
\end{equation}
\[
(0.5+ \phi X_{sp^3}) X_{\rm H} + 7 (0.5 X_{sp^2}) + 7 X_{sp^3} = 3, 
\]
which, upon re-arrangement, gives
\begin{equation}
0.5 X_{\rm H} + 7 (0.5 X_{sp^2}) + ( 7 + \phi X_{\rm H} ) X_{sp^3} = 3, 
\end{equation}
where $a = 0.5 $, $b = 3.5$ and $c = ( 7 + \phi X_{\rm H} )$. Substituting into Eqs.~(\ref{gen_R_abc}), (\ref{gen_m_abc}) and (\ref{gen_mC_abc}) we have 
\begin{equation}
R = \frac{(6X_{\rm H}-1)}{(8-13X_{\rm H})+2 \phi X_{\rm H}(1-X_{\rm H})},
\end{equation}
\begin{equation}
\bar{m} = \frac{(20-8X_{\rm H})+2 \phi X_{\rm H}(3-2X_{\rm H})}{(7+2 \phi X_{\rm H})},
\end{equation}
\begin{equation}
\bar{m_{\rm C}} =  \frac{(20-15X_{\rm H})+6 \phi X_{\rm H}(1-X_{\rm H})}{[7+2 \phi X_{\rm H}(1-X_{\rm H})]},
\end{equation}

\begin{figure}
 \resizebox{\hsize}{!}{\includegraphics{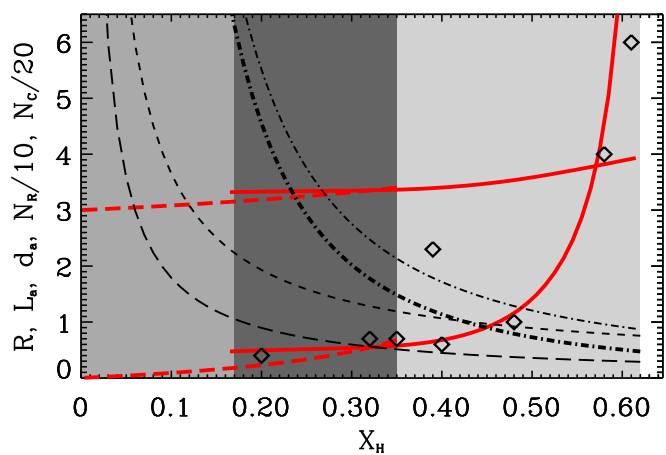}}
\caption{$R$ as a function of $X_{\rm H}$ for the best-fit RCN model (Eq.~\ref{RvsXH_fit_CH3}, lower solid red line) and DG model (with $L=2$, Eq.~\ref{DG_RvsXH}, lower dashed red line); the corresponding values of $m_{\rm C}$ for each model are shown in the upper `flatter' red lines.  
Shaded: validity-ranges for the RCN (right) and DG (left) models, darker where they overlap. 
Thick dash-dotted: mean number of rings per aromatic cluster ($N_{\rm R}/10$).
Thin dash-dotted: mean number of C atoms per aromatic cluster ($n_{\rm C}/20$).
Short dashed: aromatic domain size ($d_a$ in nm).
Long-dashed: aromatic coherence length ($L_a$ in nm).
Diamonds; laboratory data for $R$ \citep[][and references therein]{1988JVST....6.1778A}.}
\label{RvsXH_CH3}
\end{figure}

If we now include the aromatic clusters back into this model with the methyl groups included then the eRCN equations (Eqs.~\ref{atom_frac_sum_C2} and \ref{xi_Ncon_sum_C2}) can now be re-written in the form
\begin{equation}
X_{\rm H} + X_{sp^2} + X_{sp^3} = 1,
\end{equation}
\begin{equation}
0.5 X_{\rm H} + \{7 [0.5(1-f)]+Zf \} X_{sp^2} + ( 7+\phi X_{\rm H} ) X_{sp^3} = 3,
\end{equation}
where $a = 0.5 $, $b = \{7 [0.5(1-f)]+Zf \} = 0.5(7-Y_f)$ and $c = ( 7 + \phi X_{\rm H} )$. Using Eqs.~(\ref{gen_R_abc}), (\ref{gen_m_abc}) and (\ref{gen_mC_abc}) and substituting for $a, b$ and $c$ using \\
$
(c-b) = 0.5[(7+2 \phi X_{\rm H})+Y_f]
$, \\ 
$
(c-3) = 0.5(8+2 \phi X_{\rm H}) 
$, \\ 
$
(c-a) = 0.5(13+2 \phi X_{\rm H}) 
$, \\ 
$
(b-a) = 0.5(6-Y_f) 
$ and \\ 
$
(b-3) = 0.5(1-Y_f) 
$, \\ 
we obtain $R$, and the average coordination numbers, $\bar{m}$ and $\bar{m_{\rm C}}$, for our most sophisticated eRCN network, {\it i.e.}, 
\begin{equation}
R = \frac{(6X_{\rm H}-1)+Y_f(1-X_{\rm H})}{(8-13X_{\rm H})+2 \phi X_{\rm H}(1-X_{\rm H})}, 
\label{RvsXH_fit_CH3}
\end{equation} 
\begin{equation}
\bar{m} = \frac{(20-8X_{\rm H})+(Y_f+2 \phi X_{\rm H})(4-3X_{\rm H})} {(7+2 \phi X_{\rm H}+ Y_f)}, 
\end{equation}
\begin{equation}
\bar{m_{\rm C}} =  \frac{(20-15X_{\rm H})+2(2Y_f+3 \phi X_{\rm H})(1-X_{\rm H})} {(7+2 \phi X_{\rm H}+ Y_f)(1-X_{\rm H})}. 
\end{equation}
The above expression for $R$ is plotted in Figs.~\ref{RvsXH} and \ref{RvsXH_CH3}, where we see that Eq.~(\ref{RvsXH_fit_CH3}) gives a better fit to the laboratory data for the hydrogen-rich amorphous carbons. Here we have empirically assumed $\phi = 30(X_{\rm H}-0.5)^2$,  in order to give a physically-reasonable fit to the steep rise of $R$ at high $X_{\rm H}$,  and note that the fit is not particularly sensitive to the detailed form of the expression for $\phi$.  

In Fig.~\ref{RvsXH_CH3} we also plot other key quantities, namely: the number of rings per aromatic domain, $N_{\rm R} = [5.8/4.3 X_{\rm H}]^2$, the aromatic domain size, $d_a= 0.18 (3 N_{\rm R}+4)^{0.5}$~nm, and the aromatic coherence length, $L_a = 0.77/E_g({\rm eV})$\,nm $\equiv 0.77/(4.3 X_{\rm H})$\,nm (see Appendix~\ref{aromatics} for more details).

\subsubsection{A compositional de-construction of eRCNs}
\label{structural_decomp}

In this section we quantitatively examine the chemical compositional make-up and bonding configurations of eRCN-modelled a-C:Hs, {\it i.e.}, we determine their C-H$_n$ and (C-C)$_m$ bond fractions. This is done with the ultimate view of being able to predict the $X_{\rm H}$- and size-dependent, infrared spectral characteristics of these materials using the laboratory-measured cross-sections and intensities for the constituent structures \citep[{\it e.g.},][]{1998JAP....84.3836R}. The derived structures have a key bearing on the electronic and optical properties of these materials. 

In `de-constructing' the network in this way we need to consider all of the possible, fundamental and basic C$_n$H$_m$ ($n \geq 1$, $m \geq 0$) building blocks that make up solid, amorphous hydrocarbons. Table~\ref{building_blocks} presents the major bond groupings that we will consider in this study, gives their eRCN designations and the expressions that define their atomic fractions (derived later within this section). 

The eRCN model clearly cannot account for the presence of interstitial H atoms because they do not affect the network, nevertheless their effects could be important \citep[{\it e.g.},][]{1989ApPhL..54.1412S,1996MNRAS.283..343D,2011ApJ...737L..44D} and will need to be taken into account in more sophisticated models.

Note that, as above, we assume that any $sp^1$ C-H or C$\equiv$C bonds are of such low abundance that their presence within, and influence upon, the eRCN model of the a-C:H structure and composition can be ignored without loss of generality \citep[{\it e.g.},][]{1986AdPhy..35..317R,2007DiamondaRM...16.1813K}. Similarly, we also ignore the possible incorporation of oxygen and nitrogen atoms into the hydrocarbon structure. Although, the addition of $sp^1$ carbon atoms and hetero-atoms could, in principle, be incorporated into a further extension of the eRCN formalism, with the inclusion of structure-terminating and bridging groups such as \ --~C$\equiv$C~--, \ $\equiv$C--H, \ --~O~--~, \ --~OH, \ =~O, \ --~NH$_2$, \ =~NH and the substitution of N for C in aromatic rings. However, and in practice, it is difficult to do this in a physically-meaningful way because of the inherent limitations of the method (see Appendix\,\ref{extensions}). \cite{1997ApJ...476..184D} have in fact already considered the incorporation of O atoms, in the form of OH groups, into the RCN structure and find that this can lead to a strong absorption feature in the 3\,$\mu$m region. 

In their work on a-C, a-C:H and ta-C materials, with band gaps in the range $0-3$\,eV, \cite{2007DiamondaRM...16.1813K} find that their two Tauc-Lorentz (2-TL) oscillator model, for the optical properties, does not give an entirely satisfactory fit for a-C:H materials. They suggest that this is due to the presence of $sp^1$ bonds and chain-like configurations that are inconsistent with the assumptions of their model. There is now convincing analytical and experimental evidence to support this suggestion because as-deposited, hydrogen-free, cluster-assembled carbon films can exhibit a significant $sp^1$ carbon chain component, which is almost completely destroyed upon exposure to air \cite[{\it e.g.},][]{2004PhilTransRSocLondA..362.2477F}. 

As pointed out by \cite{2004PhilTransRSocLondA..362.2477F}, the nature of the clustering of the $sp^2$ phase (into chains or cage-like structures) and the orientation of those clusters can play a key role in the determination of the properties of amorphous (hydro)carbon materials. 

Using femtosecond pulsed laser ablation and deposition to form tetrahedral carbon and amorphous diamond-like films \cite{2007JAP...102g4311H,2007JChPh.126o4705H} find that the composition of these materials includes $sp^1$, $sp^2$, and $sp^3$-bonded carbon. They show that an observed excitation band in their UV-Raman spectra ($2000-2200$\,cm$^{-1}$ $\equiv 4.55-5.00\,\mu$m) is consistent with the presence of $sp^1$ chains in their materials, with $sp^1$ carbon atom fractions of $\approx 6$\%. This work clearly provides a key insight into the detailed composition and structure of tetrahedral carbon films containing $sp^2$ clusters and $sp^1$ chains. However, it yet remains to be seen whether the properties of these (hydrogen-poor) materials, often produced under ``stressful'' conditions, can be applied to interstellar carbonaceous particles formed under less-intense energy conditions and where exposure to H atoms is practically ubiquitous. 

Interest in the inclusion of nitrogen into amorphous carbons has led to the study of amorphous carbon nitrides, a-CN$_x$, in which various N-bonding configurations have been classified \citep[{\it e.g.},][]{2004PhilTransRSocLondA..362.2477F}. In their detailed study, \cite{2005A&A...432..895D} found that if nitrogen atoms are present in laboratory analogues of interstellar amorphous hydrocarbons then they tend to be found in structure-bridging groups rather than being incorporated into aromatic clusters. However, the similar frequencies for C--C and C--N modes does make the interpretation of the skeletal modes rather difficult \citep[{\it e.g.},][]{2004PhilTransRSocLondA..362.2477F}. In general, a low content or absence of `hetero-atom' groups in ISM carbonaceous materials is apparent \citep{2002ApJS..138...75P,2005A&A...432..895D} and hence the simplifying assumption to ignore them here is justified. Nevertheless, there are indications for a carbonyl component (characteristic of ketones) in the carbonaceous dust in the Seyert galaxy \object{NGC\,1068} \citep{2004A&A...423..549D}. Additionally, the incorporation of O and N into the CHON-type materials (in CHON, CH, CHO and CHN combinations) detected in the comet Halley dust \citep[{\it e.g.},][]{1987A&A...187..779C} implies that the incorporation of O and N hetero-atoms into the eRCN structure could be important for Solar System studies. However, such an extension of the eRCN model is not without its limitations, as pointed out in Appendix\,\ref{extensions}. 

\begin{table*}
\caption{Designation of, and formul\ae\ for, the major eRCN bulk material structural groups.}
\begin{center}
\begin{tabular}{cllll}
                       &                      &                  &               &                \\[-0.35cm]
\hline
\hline
                       &                      &                  &               &                     \\[-0.35cm]
  {Carbon atom}    &  {Bonding}       &  {Atomic}                 &  {Atomic fraction}  &                    \\
  {hybridisation}  &  {structure}     &  {fraction}          &                 {expression} &    {Notes}  \\[0.05cm]
\hline
                       &                      &                       &               &                    \\[-0.2cm]
                       
\multicolumn{3}{l}{\underline{Atoms}}                                                  \\[0.1cm]
H                         & ---                  &  \ \ \ \ \ \ \ \ \ $X_{\rm H}$          &  = \ \ \ $X_{\rm H}$  &                   \\
$sp^2$ C             & ---                  &  \ \ \ \ \ \ \ \ \ $X_{sp^2}$           &  = \ \ \ $X_{sp^2}$   &                   \\
$sp^3$ C             & ---                  &  \ \ \ \ \ \ \ \ \ $X_{sp^3}$           &  = \ \ \ $X_{sp^3}$   &                  \\[0.2cm]

\multicolumn{3}{l}{\underline{Aromatic C-H bonds}}                                              \\[0.1cm]

$sp^2$                 &  CH \ aromatic        &  \ \ \ \ \ \ \ \ \ $X^{2}_{\rm CH,ar}$  &  = \ \ \ $\eta \, m_{\rm ar} \ f X_{sp^2} $ & $\eta = X_{\rm H}/(1-X_{\rm H})$, the [H]/[C] atom ratio   \\
 & & & & $m_{\rm ar}$ = (aromatic cluster coord. no.)/(no. of C atoms) \\[0.2cm]

\multicolumn{3}{l}{\underline{Olefinic and aliphatic C-H bonds} - for $X_{\rm H} \leq 0.5$}                                              \\[0.1cm]

$sp^2$  &  CH \ olefinic          &  \ \ \ \ \ \ \ \ \ $X^{2}_{\rm CH}$      &  = \ \ \ $u \, \eta \, (1-f)X_{sp^2}$ &  $u = \{ 1+ ((1-m_{\rm ar})/[(1-X_{\rm H})/(f X_{sp^2}) -1 ] ) \}^{(a)} $  \\  
$sp^2$  &  CH$_2$ olefinic    &  \ \ \ \ \ \ \ \ \ $X^{2}_{\rm CH_2}$   &  = \ \ \ 0                                             &           \\

$sp^3$  &  CH \ aliphatic        &  \ \ \ \ \ \ \ \ \ $X^{3}_{\rm CH}$       &  = \ \ \ $u \, \eta \, X_{sp^3}$                   &           \\
$sp^3$  &  CH$_2$ aliphatic  &  \ \ \ \ \ \ \ \ \ $X^{3}_{\rm CH_2}$   &  = \ \ \ 0                                             &           \\
$sp^3$  &  CH$_3$ aliphatic  &  \ \ \ \ \ \ \ \ \ $X^{3}_{\rm CH_3}$   &  =  \ \ \ 0                                            &           \\[0.2cm]

\multicolumn{3}{l}{\underline{Olefinic and aliphatic C-H bonds} - for $X_{\rm H} > 0.5$}                                                  &  \\[0.1cm]

$sp^2$  &  CH \ olefinic           &  \ \ \ \ \ \ \ \ \ $X^{2}_{\rm CH}$     &  =  \ \ \ \ $\, u\, (2-\eta)\, (1-f) \, X_{sp^2}$ &    \\
$sp^2$  &  CH$_2$ olefinic     &  \ \ \ \ \ \ \ \ \ $X^{2}_{\rm CH_2}$ &  =  \ \ $2 \, u\, (\eta-1) \, (1-f) \, X_{sp^2}$   &    \\

$sp^3$  &  CH \ aliphatic         &  \ \ \ \ \ \ \ \ \ $X^{3}_{\rm CH}$     &  =  \ \ \ \ $\, u\, (2-\eta) \, X_{sp^3} / (1+\phi)$     &  $\phi = 30(X_{\rm H}-0.5)^2$  \\
$sp^3$  &  CH$_2$ aliphatic   &  \ \ \ \ \ \ \ \ \ $X^{3}_{\rm CH_2}$ &  =  \ \ $2 \, u \, (\eta-1) \, X_{sp^3} / (1+\phi)$          &    \\
$sp^3$  &  CH$_3$ aliphatic   &  \ \ \ \ \ \ \ \ \ $X^{3}_{\rm CH_3}$ &  = \ \ $3 \, u \, \phi \, X_{sp^3} \ \ \ \ \ \ \ \ \, / (1+\phi)$  &  $(1+\phi)$ re-normalises for  CH$_3$    \\[0.2cm]

\multicolumn{3}{l}{\underline{C-C bonds}$^{(b)} $}                                              \\[0.1cm]

$sp^2$                 &  C$\simeq$C aromatic &  \ \ \ \ \ \ \ \ \ $X_{\rm C\simeq C}$  &  = \ \ \ \ \ \ \ \ \ \ \ $\frac{3}{2}f \ X_{sp^2}/X_{\rm bond}$   &  $f=0.5{\rm e}^x/(1+{\rm e}^x)$   \\
$sp^2$                 &  C=C olefinic        &  \ \ \ \ \ \ \ \ \ $X_{\rm C=C}$        &  = \ \ \ $\frac{1}{2}(1-f)X_{sp^2}/X_{\rm bond}$   & $x=(X_{\rm H}-X_0)/\delta$        , $X_0 =0.35$, $\delta = 0.085$                 \\
$sp^3$                 &  C$-$C aliphatic     &  \ \ \ \ \ \ \ \ \ $X_{\rm C-C}$        &  = \ \ \ \ \ \ \ \ \ \ \ \ \ \ \ \, $X_{sp^3}/X_{\rm bond}$    &  $X_{\rm bond} = X_{\rm H} + (\frac{1}{2}+f) X_{sp^2} + X_{sp^3}$                \\[0.2cm]
\hline
\hline
                       &                      &                      &                   &  \\[-0.35cm]
\end{tabular}
\begin{list}{}{}
\item[] Notes --  (a) The parameter $u$ corrects for the fact that only the peripheral aromatic cluster carbon atoms can bond with hydrogen atoms, thus `releasing' more hydrogen atoms to bond with other non-aromatic carbon atoms.
(b) The factor $X_{\rm bond}$ in the C-C bond fractions is a required renormalisation because the total number of bonds in the eRCN $\neq$ the number of atoms, as indicated by the pre-$X$ fractions in the $sp^2$ C-C bonds expressions. 
\end{list}
\end{center}
\label{building_blocks}
\end{table*}

We note that the expressions for the C-C bonds in Table~\ref{building_blocks} follow trivially from the earlier-defined expressions for the $sp^2$ and $sp^3$ carbon atom content of the eRCN ({\it e.g.}, see \S~\ref{RCN_basics}). For the more abundant C-H bond configurations we adopt a simple statistical approach to arrive at the C-H bond definitions, {\it i.e.}, we assume an equipartition of the hydrogen atoms between the $sp^2$ and $sp^3$ carbon atom sites where the H atoms can bond. Based on this assumption we derive expressions for the aromatic CH, olefinic CH$_n$ ($n = 1,2$) and aliphatic CH$_n$ ($n = 1,2,3$) bond concentrations.

For (e)RCNs the case where $X_{\rm H} = 0.5$ provides something of a critical point, which corresponds to structures with $({\rm H}/{\rm C}) = 1$, {\it i.e.}, where, statistically, each carbon atom is associated with one hydrogen atom. We therefore now consider the two regimes, $X_{\rm H} \leq 0.5$ and $X_{\rm H} > 0.5$, in our de-construction of the eRCN bonding. In both cases the hydrogen-to-carbon atom concentration ratio can be expressed as,
\begin{equation}
\frac{[H]}{[C]} = \frac{X_{\rm H}}{(1-f)X_{sp^2}+X_{sp^3}} =  \frac{X_{\rm H}}{(1-X_{\rm H})-fX_{sp^2}} \equiv \eta, 
\end{equation}
where the denominator is the olefinic $sp^2$ and aliphatic $sp^3$ carbon atom fraction. As can be seen, we have for the moment, ignored the aromatic carbon and its associated hydrogen content, as indicated by the $-fX_{sp^2}$ term in the denominator. We return to this missing aromatic carbon component later in this section. 

\begin{figure}
 \resizebox{\hsize}{!}{\includegraphics{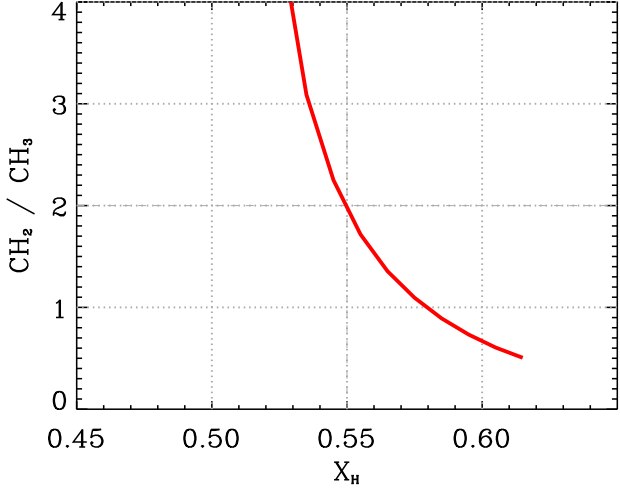}}
 \caption{ The CH$_2$/CH$_3$ abundance ratio as a function of $X_{\rm H}$ for the aliphatic carbon component  in the best-fit eRCN model (Eq.~\ref{RvsXH_fit_CH3}). }
 \label{fig_decomp_4}
 \end{figure}

\begin{figure}
 \resizebox{\hsize}{!}{\includegraphics{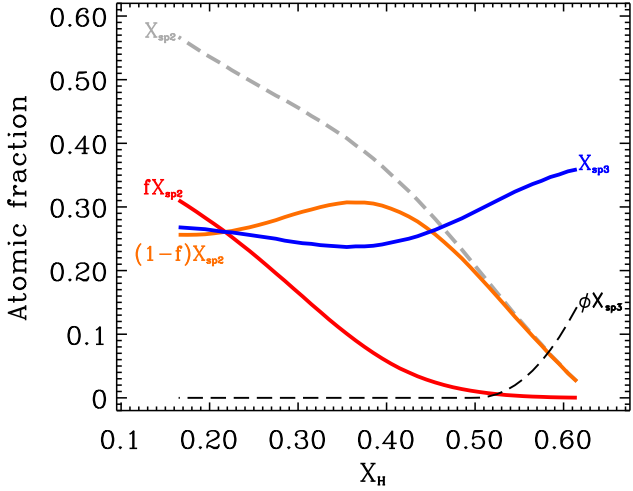}}
 \caption{The aromatic, $fX_{sp2}$ (red), olefinic, $(1-f)X_{sp2}$ (orange), total $sp^2$, $X_{sp2}$ (grey dashed), aliphatic, $X_{sp3}$ (blue) and CH$_3$, $\phi X_{sp3}$ (black dashed), carbon atom fractions as a function of $X_{\rm H}$ for the best-fit eRCN model (Eq.~\ref{RvsXH_fit_CH3}).}
 \label{fig_decomp_1}
\end{figure}

\begin{figure}
 \resizebox{\hsize}{!}{\includegraphics{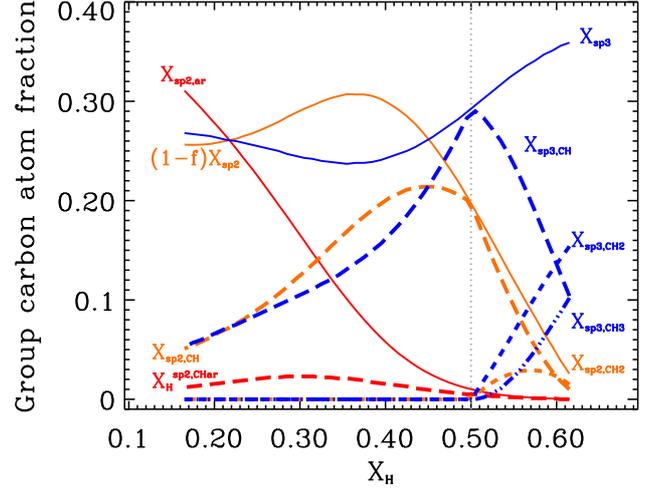}}
 \caption{As for Fig.~\ref{fig_decomp_1} with the aromatic ($X_{\rm H}^{sp^2,{\rm CHar}}$, red dashed), olefinic ($X_{\rm H}^{sp^2,{\rm CH}}$, $X_{\rm H}^{sp^2,{\rm CH_2}}$, orange dashed) and aliphatic ($X_{\rm H}^{sp^3,{\rm CH}}$, $X_{\rm H}^{sp^3,{\rm CH_2}}$, blue dashed; $X_{\rm H}^{sp^3,{\rm CH_3}}$, blue triple dot-dashed) CH$_n$ fractions also shown, where in most cases the H subscripts have been dropped and the superscripts written as subscripts for clarity.}
 \label{fig_decomp_2}
\end{figure}

\begin{figure}
 \resizebox{\hsize}{!}{\includegraphics{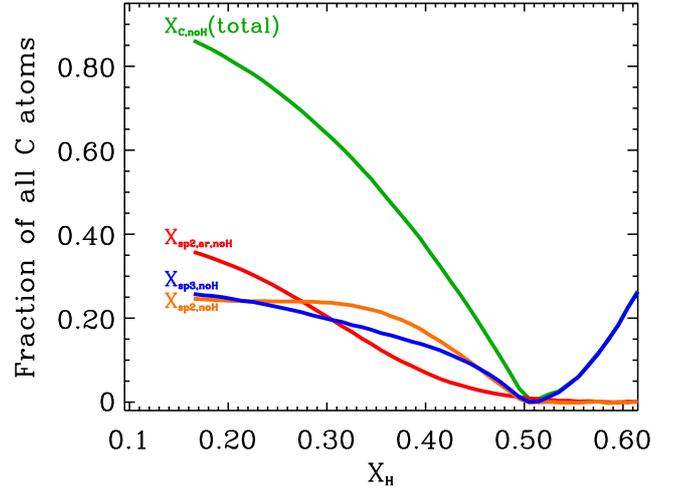}}
 \caption{The fraction of all carbon atoms ($X_{\rm C,noH}$(total), green), $sp^2$ aromatic ($X_{sp^2,{\rm ar,noH}}$, red),  olefinic ($X_{sp^2,{\rm noH}}$, orange) and $sp^3$ ($X_{sp^3,{\rm noH}}$, blue) carbon atoms  as a function of $X_{\rm H}$ that are not attached to a hydrogen atom, for the  best-fit eRCN model (Eq.~\ref{RvsXH_fit_CH3}).}
 \label{fig_decomp_3}
\end{figure}

Fig.~\ref{fig_decomp_4} shows the CH$_2$/CH$_3$ abundance ratio as a function of $X_{\rm H}$ for the aliphatic carbon component in the best-fit eRCN model. In the diffuse ISM this ratio appears to be $2.5\pm0.4$ \citep{2002ApJS..138...75P} and in laboratory a-C:H is close to 2 \citep{2005A&A...432..895D}, which would seem to imply, based on the eRCN model, that the large carbonaceous grains in the ISM, that are observed in absorption, must be rather H-rich with $X_{\rm H}$ lying in a narrow range close to $\sim 0.53-0.55$ and a H/C ratio of the order of 1.2.

For $X_{\rm H} \leq 0.5$ each carbon atom is, at most, singly hydrogenated (see above) and we can ignore CH$_2$ and CH$_3$ groups. We then have
\begin{equation}
X^i_{\rm CH} =  \frac{u \, X_{\rm H} \, X_i}{(1-X_{\rm H})-fX_{sp^2}}  = u \, \eta \, X_i, 
\end{equation}
where $i=2$ ($i=3$) corresponds to single $sp^2$ olefinic CH ($sp^3$ aliphatic CH) bonds, and $u$ is an aromatic C-H abundance re-normalising factor (see Table~\ref{building_blocks}). $X_i$ represents the corresponding carbon atom fraction, {\it {\it i.e.}},  $(1-f)X_{sp^2}$ or $X_{sp^3}$. Particularly important in the case of these hydrogen-poor eRCNs is the need to consider the $sp^2$ (polycyclic) aromatic cluster hydrogen content. From Table~\ref{PAH_cluster_params} (see also Fig.~\ref{fig_m_per_c_atom}) we can see that the aromatic cluster coordination number per carbon atom for an aromatic cluster is given by the cluster coordination number, $m_{\rm cluster}$, divided by the number of constituent carbon atoms, $n_{\rm C}$,
\begin{equation}
\frac{m_{\rm cluster}}{n_{\rm C}} = \frac{3.5 \surd N_{\rm R}+2.5}{2N_{\rm R}+3.5\surd N_{\rm R}+0.5} \equiv m_{\rm ar}, 
\end{equation}
where we have here assumed that the aromatic clusters are compact. Fig.~\ref{fig_m_per_c_atom} shows the behaviour of $m_{\rm ar}$ as a function of the number of aromatic rings, $N_{\rm R}$, in the cluster. Appendix~\ref{aromatics} gives the details of the aromatic cluster characterising parameters. For the aromatic C-H fraction we then have
\begin{equation}
X^2_{\rm CH,ar} =  \eta \, m_{\rm ar} \, f X_{sp^2}. 
\end{equation}

For $X_{\rm H} > 0.5$, a carbon atom can be multiply hydrogenated and we therefore need to calculate the concentrations of CH$_n$ groups for $sp^2$ (where $i=2$ and $n = 1$ or 2) and $sp^3$ (where $i=3$ and $n = 1$, 2 or 3). In this case it can be shown that, for $n = 1$ and 2, 
\begin{equation}
X^i_{{\rm CH}_n} =  u \, n \, [(-1)^n (n \eta-2)] \, X_i \, \left[ \ +0_{\ (i=2)}, \  -\frac{1}{3} X^3_{{\rm CH}_3, \ (i=3)} \ \right],
\end{equation}
where the term in the left hand square brackets gives $(2-\eta)$, for CH groups, or $2(\eta-1)$, for CH$_2$ groups (see Table~\ref{building_blocks}), and is a C-H bond abundance factor.
For the fractional abundance of methly groups, {\it i.e.,} $n=3$ (see \S~\ref{inc_methyl}), we have
\begin{equation}
X^3_{{\rm CH}_3} =  3 \,u \, \phi \, X_{sp^3}.
\end{equation}

Figs.~\ref{fig_decomp_1} and \ref{fig_decomp_2} show the carbon atom and CH$_n$ bond fractions, derived using the expressions (indicated in full in Table~\ref{building_blocks}), as a function of $X_{\rm H}$ for the best-fit eRCN model (Eq.~\ref{RvsXH_fit_CH3}). The general trends to be seen in these figures are that, with decreasing $X_{\rm H}$, H-rich aliphatics give way to H-poorer olefinics and aromatics, and that $sp^3$ CH$_3$ and CH$_2$ groups are progressively replaced by CH$_2$ and then CH olefinics. Note that the aromatic CH abundance is rather low in these materials, which even in their H-poorest state contain more, and about equal fractions of, aliphatic and olefinic CH groups. 

In Fig.~\ref{fig_decomp_3} we show the fractions of the eRCN constituent carbon atoms that are not bonded to a hydrogen atom. The indicated trends are rather similar with the, not-unexpected, exception that for low values of $X_{\rm H}$ most of the aromatic $sp^2$ carbon atoms are not hydrogenated, as can clearly be seen by comparison with Fig.~\ref{fig_decomp_2}.

\subsubsection{The predicted infrared spectra of bulk eRCN materials}
\label{sect_eRCN_spectra}

\begin{figure} 
 \resizebox{\hsize}{!}{\includegraphics{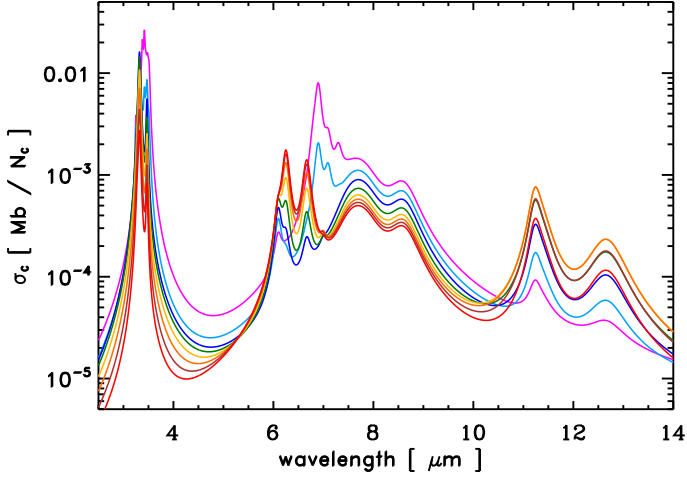}}
 \caption{The adopted wavelength-dependent, IR band cross-sections per carbon atom, $\sigma_{\rm C}$ (1\,Mb $= 10^{-18}$\,cm$^2$), for the eRCN model as a function of $X_{\rm H}$ ($\equiv E_{\rm g}$). The upper curves (purple) are for the H-rich, wide band gap materials and the cross-sections, in the 8\,$\mu$m region, decrease with decreasing $X_{\rm H}$ from top to bottom  -- $X_{\rm H} = 0.58$ (purple), 0.52 (cobalt), 0.46 (blue), 0.40 (green), 0.34 (yellow), 0.29 (orange), 0.23 (brown) and 0.17 (red).} 
 \label{fig_sigma_C}
\end{figure}

\begin{figure}
 \resizebox{\hsize}{!}{\includegraphics{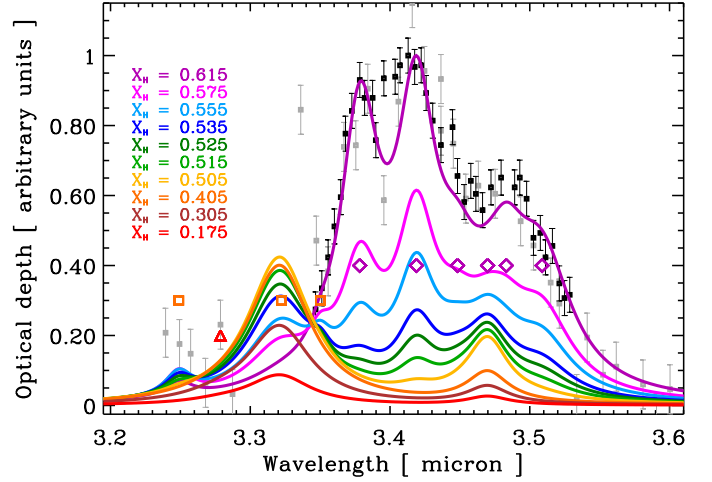}}
 \caption{The predicted eRCN spectrum in the $3.2-3.6$\,$\mu$m C-H stretching region as a function of $X_{\rm H}$ calculated using the structural de-composition described in \S\ref{structural_decomp} and the data in Table~\ref{spectral_bands}. The diamonds, squares and triangles indicate the aliphatic, olefinic and aromatic band positions, respectively (see Table~\ref{spectral_bands}).  The data with error bars are for the diffuse ISM absorption along the Galactic Centre \object{IRS6E} line of sight (black) and \object{Cyg OB2 No. 12} (grey) \citep[taken from][]{2002ApJS..138...75P}.}
 \label{stretching_spectrum}
\end{figure}

\begin{figure}
 \resizebox{\hsize}{!}{\includegraphics{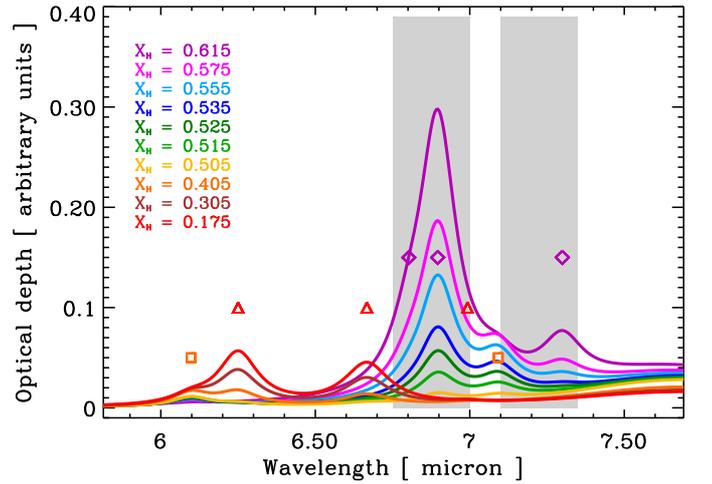}}
 \caption{The predicted eRCN spectrum in the $6.0-7.5$\,$\mu$m C-H bending and C-C mode region as a function of $X_{\rm H}$ calculated using the structural de-composition described in \S\ref{structural_decomp} and the data in Table~\ref{spectral_bands}. The symbols are the same as for the previous figure. The grey shaded areas indicate the approximate full width of the observed a-C:H bands in this region \citep[{\it e.g.},][]{2002ApJS..138...75P,2005A&A...432..895D}.}
 \label{bending_spectrum}
\end{figure}

Adopting a compositional structure and the appropriate band parameters allows a prediction of the IR spectra, as was done for HAC materials in the 3\,$\mu$m region \citep[{\it e.g.},][]{1995ApJ...445..240D,1997ApJ...476..184D,2005ApJ...626..933D}. This kind of study then enables a search for RCN solutions that would be compatible with observations \citep[{\it e.g.},][]{1995ApJ...445..240D}.

\begin{figure*}
\vspace{-5.0cm}
 \resizebox{\hsize}{!}{\includegraphics{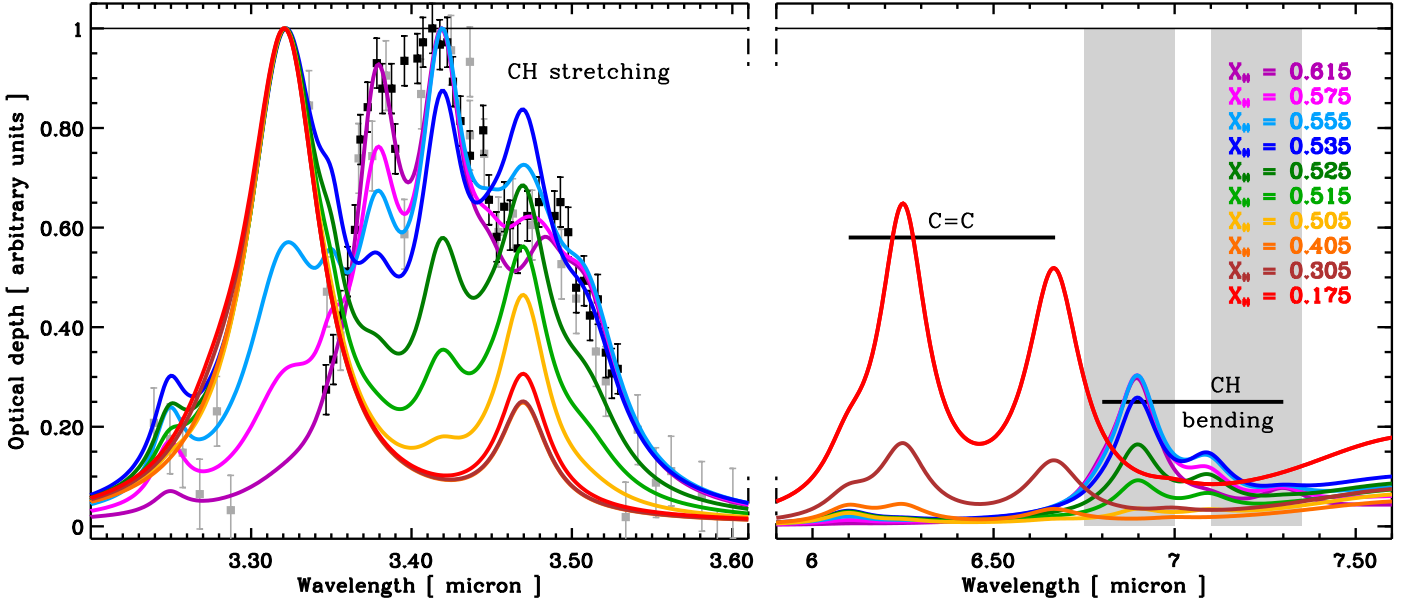}}
 \caption{The predicted full eRCN spectrum in the $3-8 \mu$m C-H stretching and bending and C-C mode region as a function of $X_{\rm H}$ calculated using the structural de-composition described in \S\ref{structural_decomp} and the data in Table~\ref{spectral_bands}. Here the spectra are normalised to the strongest band for each value of $X_{\rm H}$. The data with error bars and the shaded areas indicate the diffuse ISM bands as per Figs.~\ref{stretching_spectrum} and \ref{bending_spectrum}.}
 \label{full_spectrum_eRCN}
\end{figure*}

Using typical C-H and C-C band positions, widths and cross-sections taken from the literature (see Table~\ref{spectral_bands} and references therein, and Fig.~\ref{fig_sigma_C}) and the eRCN structural de-construction described in \S\ref{structural_decomp} and Table~\ref{building_blocks}, we show the predicted optical depth spectra $\tau(\nu)$, in arbitrary units, for bulk eRCN materials in the 3\,$\mu$m and $6-7$\,$\mu$m regions in Figs.~\ref{stretching_spectrum} and \ref{bending_spectrum}, respectively. Here the optical depth normalised by the C atom column density, $N_{\rm C}$, is given by
\begin{equation}
\frac{\tau(\nu)}{N_{\rm C}} = \sum_i \frac{\sigma_{{\rm C},i} \ X^h_i \ g_i(\nu)}{\int_0^\infty g_i(\nu) \ d \nu },
\label{ eq_tau_definition}
\end{equation}
where $\sigma_{{\rm C},i}$ is the integrated cross-section per C atom for band $i$ (see Table~\ref{spectral_bands} and Fig.~\ref{fig_sigma_C}), $X^h_i$ is the relative abundance (atomic fraction) for group $i$ (C$_n$H$_m$, where $n=1,2$ and $m=0,1,2,3$) of C atom hybridisation state $h$ (where $2 \equiv sp^2$ or $3 \equiv sp^3$, in keeping with the nomenclature used in Tables~\ref{building_blocks} and \ref{spectral_bands}) and $g(\nu)$ is the band profile. Anticipating the required behaviour of these bands at long wavelengths (FIR to cm), we here assume intensity-normalised, Drude profiles for all of the bands, {\it i.e.}, 
\begin{equation}
g(\nu) = \frac{(\gamma \nu)^2}{(\nu^2-\nu_0^2)^2 + (\gamma \nu)^2},
\label{eq_Drude}
\end{equation}
where $\nu$ is in wavenumbers, $\nu_0$ is the band centre and $\gamma$ the band width. 
The spectra in Figs.~\ref{stretching_spectrum} and \ref{bending_spectrum} have been normalised to the peak of the strongest band exhibited for all of the displayed, $X_{\rm H}$-dependent compositional spectra ({\it i.e.}, the 3.42 
band in the $X_{\rm H} = 0.615$ sample). In Fig.~\ref{full_spectrum_eRCN} we show the same spectral data but normalised to the strongest band in each spectrum in order to better indicate the evolution of the form of the spectra as a function of $X_{\rm H}$. 

%
\begin{figure*}
\vspace{-5.0cm}
 \resizebox{\hsize}{!}{\includegraphics{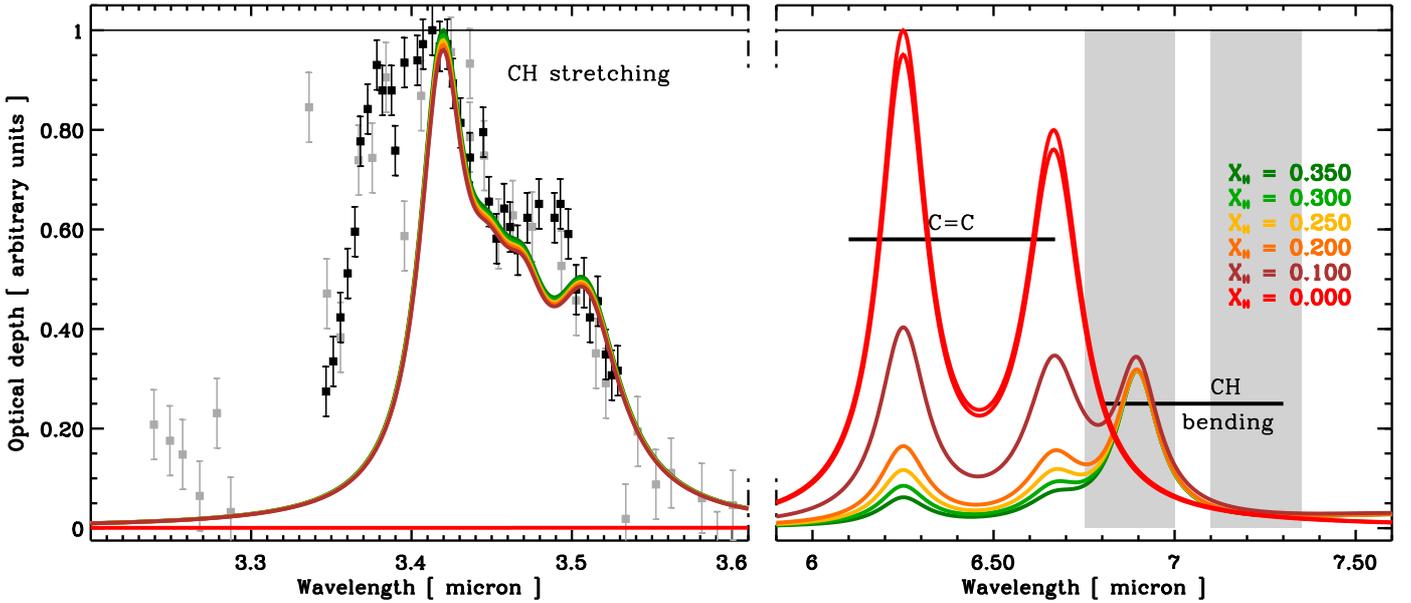}}
 \caption{The predicted full DG spectrum in the $3-8 \mu$m C-H stretching and bending and C-C mode region as a function of $X_{\rm H}$ calculated using the structural de-composition described in \S\ref{structural_decomp} and the data in Table~\ref{spectral_bands}. Here the spectra are normalised to the strongest band for each value of $X_{\rm H}$. On the left the strongest bands are for high $X_{\rm H} > 0$ (green to brown), while the opposite is true for the longer wavelength bands on the right (shortward of 6.8\,$\mu$m).  The data with error bars and the shaded areas indicate the diffuse ISM bands as per Figs.~\ref{stretching_spectrum} and \ref{bending_spectrum}.}
 \label{full_spectrum_DG}
\end{figure*}

\begin{table*}
\caption{The adopted C-H and C-C band modes in eRCN bulk materials: 
band centre ($\nu_0$), width ($\delta$) and integrated cross-section ($\sigma$).}
\begin{center}
\begin{tabular}{cccclll}
                       &              &               &               &               &                \\[-0.35cm]
\hline
\hline
                       &                       &              &                &               &                     \\[-0.35cm]
                      &  $\nu_0$                            &  $\delta$           &  $\sigma$         &         Band       &             eRCN            &                    \\
         no.        &  [ cm$^{-1}$ ( $\mu$m ) ]   &  [ cm$^{-1}$ ]   &   [ $\times 10^{-18}$ cm$^{2}$ ]   &        assignment         &   designation  & Notes  \\[0.05cm]
\hline
                       &                      &            &           &               &                    \\[-0.2cm]
                       
 & \multicolumn{6}{l}{\underline{C-H stretching modes}}                                  \\[0.1cm]
       1    &    3078  ( 3.25 )      &    22.5    &    1.40   &    $sp^2$ CH$_2$  olefinic,  asy.     &    X$^2_{\rm CH_2}$            &    \\ 
       2    &    3050  ( 3.28 )      &    53.1    &    1.50   &    $sp^2$ CH          aromatic            &    X$^2_{\rm CH_{\rm ar}}$  &    \\
       3    &    3010  ( 3.32 )      &    47.1    &    2.50   &    $sp^2$ CH          olefinic              &    X$^2_{\rm CH}$                 &    \\
       4    &    2985  ( 3.35 )      &    17.7    &    1.15   &    $sp^2$ CH$_2$  olefinic,  sym.    &    X$^2_{\rm CH_2}$             &    \\
       5    &    2960  ( 3.38 )      &    29.3    &    5.00   &    $sp^3$ CH$_3$  aliphatic, asy.    &    $\phi$ X$^3_{\rm CH_3}$  &    \\
       6    &    2925  ( 3.42 )      &    28.9    &    3.30   &    $sp^3$ CH$_2$  aliphatic, asy.    &    X$^3_{\rm CH_2}$             &    \\
       7    &    2900  ( 3.45 )      &    30.0    &    0.50   &    $sp^3$ CH$_2$  aliphatic            &    X$^3_{\rm CH}$                &    \\
       8    &    2882  ( 3.47 )      &    30.2    &    1.00   &    $sp^3$ CH tertiary  aliphatic        &    X$^3_{\rm CH_2}$             &    \\
       9    &    2871  ( 3.48 )      &    27.8    &    1.45   &    $sp^3$ CH$_3$  aliphatic, sym.    &    $\phi$ X$^3_{\rm CH_3}$  &    \\
     10    &    2850  ( 3.51 )      &    41.8    &    2.20   &    $sp^3$ CH$_2$  aliphatic, sym.    &    X$^3_{\rm CH_2}$             &    \\[0.2cm]
                       
 & \multicolumn{6}{l}{\underline{C-H bending modes}}                                     \\[0.1cm]
      11    &    1470  ( 6.80 )    &    30.0    &    0.30     &    $sp^3$ CH$_3$  aliphatic, asy.             &    $\phi$ X$^3_{\rm CH_3}$    &    est. width  \\
      12    &    1450  ( 6.90 )    &    30.0    &    1.20     &    $sp^3$ CH$_2$  aliphatic                     &    X$^3_{\rm CH_2}$              &    est. width  \\
      13    &    1430  ( 6.99 )    &    30.0    &    0.10     &    $sp^2$ CH          aromatic                    &     X$^2_{\rm CH_{\rm ar}}$    &    est. width  \\
      14    &    1410  ( 7.09 )    &    30.0    &    1.00     &    $sp^2$ CH$_2$  olefinic                       &    X$^2_{\rm CH_2}$              &    est. width  \\
      15    &    1400  ( 7.14 )    &    30.0    &    0.10     &    $sp^3$ (CH$_3$)$_3$ aliphatic, sym.   &     X$^3_{\rm (CH_3)_3}$       &    not included  \\
      16    &    1370  ( 7.30 )    &    30.0    &    0.30     &    $sp^3$ CH$_3$   aliphatic, sym.           &   $\phi$ X$^3_{\rm CH_3}$     &    est. width  \\[0.2cm]
                       
 & \multicolumn{6}{l}{\underline{C-C modes}}                                                  \\[0.1cm]
      17    &    1600  ( 6.25 )    &    40.0  &    0.19      &    $sp^2$ C$\simeq$C   aromatic           &    X$_{\rm C\simeq C}$    &    est. width    \\
      18    &    1500  ( 6.67 )    &    40.0  &    0.15      &    $sp^2$ C$\simeq$C   aromatic           &    X$_{\rm C\simeq C}$    &    est. width    \\
      19    &    1640  ( 6.10 )    &    40.0  &    0.10      &    $sp^2$ C=C   olefinic                          &    X$_{\rm C=C}$              &    all data est. \\[0.2cm]
                             
 & \multicolumn{6}{l}{\underline{additionally-assumed and estimated C-H modes}}  \\[0.1cm]
      20    &    3020  (   3.31 )    &    50.0  &    0.50      &    $sp^2$ CH   olefinic                          &    X$^2_{\rm CH}$                &    est. intensity \& width    \\
      21    &      890  ( 11.24 )    &    20.0  &    0.50      &    $sp^2$ CH   aromatic                       &    X$^2_{\rm CH_{\rm ar}}$   &    est. intensity \& width    \\
      22    &      880  ( 11.36 )    &    40.0  &    0.50      &    $sp^2$ CH   aromatic                       &    X$^2_{\rm CH_{\rm ar}}$   &    est. intensity \& width    \\
      23    &      790  ( 12.66 )    &    50.0  &    0.50      &    $sp^2$ CH   aromatic                       &    X$^2_{\rm CH_{\rm ar}}$   &    est. intensity \& width    \\[0.2cm]
                             
 & \multicolumn{6}{l}{\underline{additionally-assumed and estimated C-C modes}}  \\[0.1cm]
      24    &    1328  (  7.53  )    &  120.0  &    0.10      &    $sp^3$ C-C  aliphatic                        &    X$^3_{\rm C-C}$               &    est. intensity \& width    \\
      25    &    1300  (  7.69 )     &  120.0  &    0.10      &    $sp^3$ C-C  aliphatic                        &    X$^3_{\rm C-C}$               &    est. intensity \& width    \\
      26    &    1274  (  7.85  )    &  120.0  &    0.10      &    $sp^3$ C-C  aliphatic                        &    X$^3_{\rm C-C}$               &    est. intensity \& width    \\
      27    &    1163  (  8.60  )    &    90.0  &    0.10      &    $sp^3$ C-C  aliphatic                        &    X$^3_{\rm C-C}$               &    est. intensity \& width    \\[0.2cm]

\hline
\hline
                       &                      &                      &        &           &  \\[-0.35cm]
\end{tabular}
\begin{list}{}{}
\item[] Notes --  The data, except where indicated as estimated (est.), are taken from \cite{2004A&A...423..549D,2004A&A...423L..33D,1994A&A...281..923J,1998JAP....84.3836R,1967ApSRv...1...29W}
and E. Dartois private communication. sym. and asy. indicate symmetric and asymmetric modes, respectively. The 3.47\,$\mu$m band is estimated from \cite{2008ApJ...682L.101M} and assumed to be similar to the CH$_2$ group properties. The 3.38\,$\mu$m CH$_3$ band intensity has been scaled up by a factor of 1.4, with respect to the expected value, in order to get a better fit to the the Galactic Centre data (Figs.~\ref{stretching_spectrum} and \ref{full_spectrum_eRCN}). Data for the additionally-assumed bands are taken from \cite{1986AdPhy..35..317R} and \cite{2001A&A...372..981V}; here we have assumed cross-sections of $5 \times 10^{-19}$\,cm$^2$ and $10^{-19}$\,cm$^2$ for all CH and CC modes, respectively. 
\end{list}
\end{center}
\label{spectral_bands}
\end{table*}

From Figs.~\ref{stretching_spectrum}, \ref{bending_spectrum} and \ref{full_spectrum_eRCN} it can clearly be seen that, as $X_{\rm H}$ decreases, the band intensities weaken with the transition from aliphatic-dominated structures to olefinic and aromatic-rich materials. It is interesting to note, from a close look at Figs.~\ref{stretching_spectrum} and \ref{bending_spectrum}, that there is a rather clear separation of the spectral properties of eRCNs at $X_{\rm H} \sim 0.5$ as the dominant structure changes from aliphatic to olefinic/aromatic. Indeed the structure and spectra of eRCNs evolve `rapidly'  in the region  $0.5 < X_{\rm H} < 0.62$ ({\it c.f.,} the steepness of $R$ in this region of $X_{\rm H}$ in Fig.~\ref{RvsXH}), as a result of the loss of aliphatic material, and then evolve more `slowly' for $X_{\rm H} < 0.5$, where the transition is principally from olefinic to aromatic composition as the $sp^2$ regions `coalesce' into larger aromatic domains as $X_{\rm H}$ decreases. Thus, and as expected from the cross-sections indicated in Table~\ref{spectral_bands}, at low $X_{\rm H}$ the spectra show only the relatively weaker aromatic bands occurring at shorter wavelengths than for the aliphatic, H-rich eRCNs ($X_{\rm H} > 0.5$) in both of the spectral windows shown in Figs.~\ref{stretching_spectrum}, \ref{bending_spectrum} and \ref{full_spectrum_eRCN}.

\subsection{The defective graphite (DG) model for H-poor amorphous hydrocarbons}
\label{sect_DG_model}

As shown above the eRCN model is only valid for amorphous hydrocarbons with $X_H \gtrsim 0.2$. For H-poor amorphous hydrocarbons we therefore need to develope a new approach to modelling their structural properties. In order to do this we adopt and modify the `defective graphite' (DG) formalism of \cite{1990JAP....67.1007T}, which considers carbon atom vacancies in an otherwise perfect graphite lattice. Such vacancies leave the three adjacent carbon atoms, which are supposed to transform from $sp^2$ to $sp^3$, each with two dangling bonds because the contiguity of the aromatic structure has been broken. Essentially, this lattice vacancy approach is a percolation issue because, with increasing vacancy concentration, a percolation threshold is reached, which is the point where the aromatic domains cease to percolate and become isolated \cite[{\it e.g.},][]{1994PercolationTheoryBook}. Ultimately all aromatic character will be lost at some higher vacancy concentration. 

\begin{figure}
 \resizebox{\hsize}{!}{\includegraphics{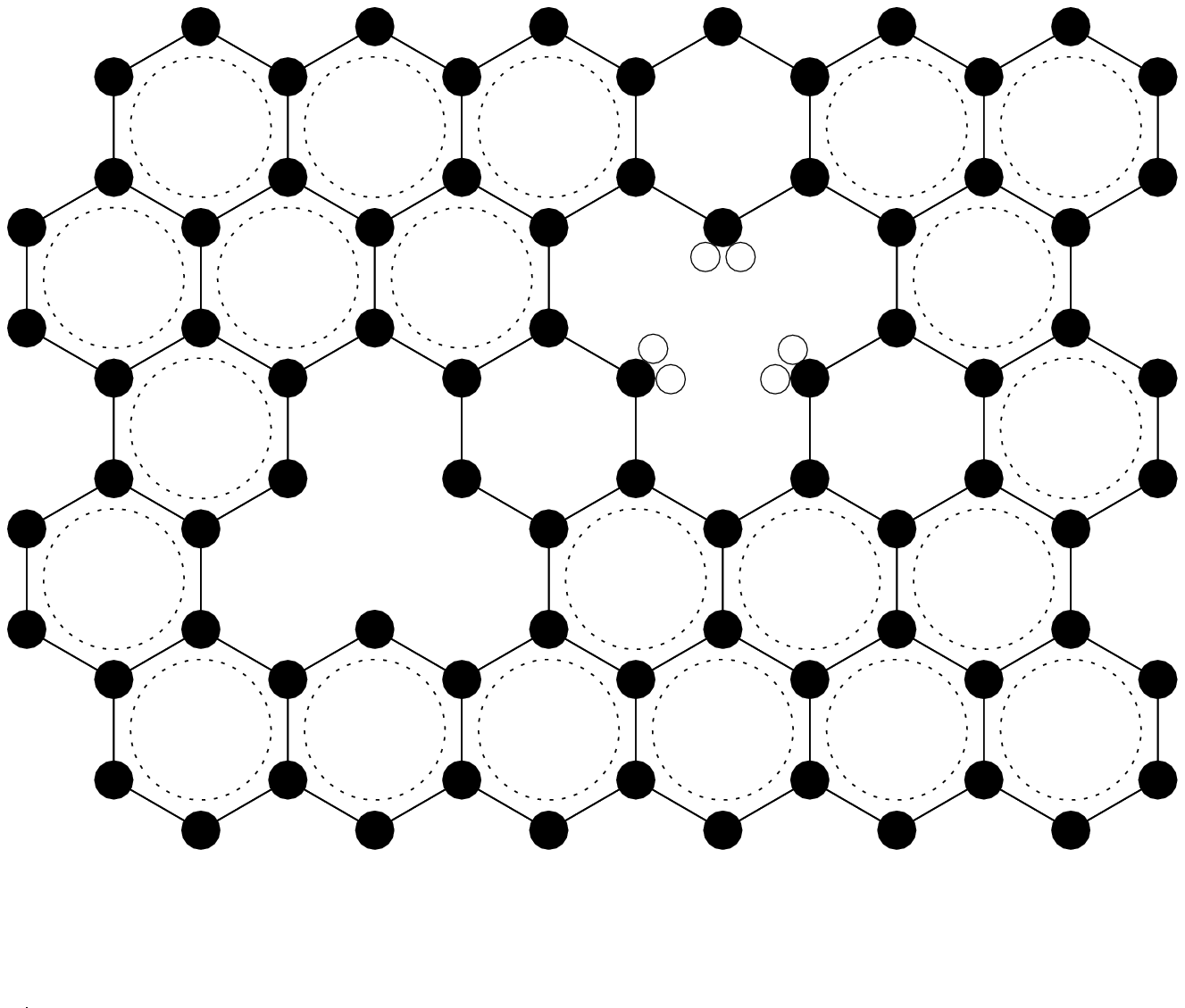}}
 \caption{The defected graphite (DG) network model. The filled (open) circles represent carbon (hydrogen) atoms in an aromatic network. The inset dotted circles indicate rings with aromatic character.}
\label{DG_defects}
\end{figure}

Fig.~\ref{DG_defects} shows a graphite basal plane with two carbon atom vacancies, and indicates the essential elements of the DG network model. In Fig.~\ref{DG_defects} the lower, left defect contains three radical carbon atoms and results in a loss of the aromatic character in the three concerned rings.  Each carbon atom adjacent to a defect (carbon atom vacancy) can then be `passivated' by the addition of two hydrogen atoms, to form three associated $sp^3$ CH$_2$ groups within the graphite basal plane. The loss of one carbon atom is then compensated by the addition of six hydrogen atoms. In the upper right defect the three adjacent carbon atoms have been transformed to $sp^3$ and each has been bonded to two hydrogen atoms. In this case there is an associated loss of the aromatic character in six adjacent rings.

If we now consider a defect-free, graphite lattice with $N_{\rm C}$ carbon atoms, into which we introduce $N_{\rm v}$ carbon atom vacancies (defects) each of which is associated with six hydrogen atoms, then
\begin{equation}
X_{\rm H}= \frac{N_{\rm H}}{N_{\rm C} + N_{\rm H}} = \frac{6 N_{\rm v}}{(N_{\rm C} - N_{\rm v}) + 6 N_{\rm v}}. 
\end{equation}
If we now define the vacancy fraction within the defective graphite lattice as $Q = (N_{\rm v}/N_{\rm C})$  then
\begin{equation}
X_{\rm H}= \frac{6Q}{1+5Q}, 
\label{DG_XH}
\end{equation}
where the total number of carbon and hydrogen atoms in the lattice is $(1+5Q)$. The number of $sp^3$ sites is then $3 N_{\rm v}$ and their atomic fraction is $3(N_{\rm v}/N_{\rm C}) \equiv 3Q$. Similarly, the atomic fraction of $sp^2$ sites is $1-(N_{\rm v}/N_{\rm C}) - 3Q = 1-4Q$ and for the DG model we can write
\begin{equation}
X_{sp^3} = \frac{3Q}{1+5Q}, 
\end{equation}
\begin{equation}
X_{sp^2} = \frac{1-4Q}{1+5Q}, 
\end{equation}
\begin{equation}
R = \frac{X_{sp^3}}{X_{sp^2}} =  \frac{3Q}{1-4Q}.
\end{equation}
Substituting for $Q$ using Eq.~(\ref{DG_XH}), {\it i.e.}, $Q = X_{\rm H}/(6-5X_{\rm H})$, we then have
\begin{equation}
R = \frac{3X_{\rm H}/(6-5X_{\rm H})}{1-4X_{\rm H}/(6-5X_{\rm H})} = \frac{3X_{\rm H}}{(6-9X_{\rm H})},
\end{equation}
which indicates that the range of validity for the DG model is $0 \leq X_{\rm H} \leq 0.66$. However, the upper limit here is really determined by the defect concentration and the defected graphite metal-to-insulator transition. \cite{1990JAP....67.1007T} determine that this key transition occurs at $Q = 1/24 \ (\equiv 0.042)$ or, equivalently (via substitution into Eq.~\ref{DG_XH}), for $X_{\rm H} = 0.21$. We note that for $Q > 1/12$ ($X_{\rm H} > 0.35$) the structure is no longer a contiguous network and so $X_{\rm H} = 0.35$ ({\it i.e.}, a composition of $\approx$~CH$_{0.3}$ or C$_3$H) represents a `hard' upper limit to $X_{\rm H}$ and the validity of the DG model.  

From the point of view of percolation, a hexagonal (honeycomb) lattice has a percolation threshold $p_c = 0.6962$, which is the minimum site concentration for an infinite network in an infinite lattice \citep[{\it e.g.},][]{1994PercolationTheoryBook}. The critical vacancy threshold is then $(1-p_c) = 0.3038$. However, in this infinite honeycomb network not all sites are part of hexagonal ring systems and so this is not an appropriate model for a contiguous aromatic lattice, where each 6-fold  aromatic ring is made up of six carbon atoms each shared by three rings ($=6 \times \frac{1}{3}$, {\it i.e.}, two carbon atoms per ring). Given that, and that each vacancy leads to the de-aromatisation of three rings (with no H atom addition), the critical vacancy threshold for a contiguous aromatic network is then $(1-p_c)/(2\times3) = 0.051$, which is very close to the critical defect concentration of 0.042 for the metal-to-insulator transition in graphite found by \cite{1990JAP....67.1007T}. Thus, for vacancy concentrations larger than $\simeq 0.04$ the percolation threshold is reached and the aromatic domains no longer percolate but become isolated from one another. 

A lower limit to the H atom content of amorphous hydrocarbons comes from the results of irradiation studies, which show that H atom loss from a-C:H stops at about $X_{\rm H} \approx 0.05$ \citep[{\it e.g.},][]{1989JAP....66.3248A,1996MCP...46...198M,2011A&A...529A.146G}. Such a lower limit to the DG model validity, and as an indicator of the minimum possible H atom content in processed a-C:H materials, could have important consequences for the evolution of dust under ISM conditions if this same lower limit holds true there. For a further discussion on this point see paper~II (\S\,5.2). 

In the same way as for the eRCN model we can derive the average atomic coordination number, $m$, and average carbon atom coordination number, $m_{\rm C}$, which are
\begin{equation}
{\bar m} = X_{\rm H}+3X_{sp^2}+4X_{sp^3} = \frac{1}{2} (6-3X_{\rm H}), 
\label{DG_coord}
\end{equation}
\begin{equation}
{\bar m_{\rm C}} =  3\frac{X_{sp^2}}{X_{sp^2}+X_{sp^3}} +4\frac{X_{sp^3}}{X_{sp^2}+X_{sp^3}} = \frac{1}{2}  \frac{(6-5X_{\rm H})}{(1-X_{\rm H})}.
\label{DG_C_coord}
\end{equation}
The fraction of $sp^2$ atoms in aromatic rings is equal to $X_{sp^2}$ in the DG case because all of them are part of a ring, therefore substituting for $Q = X_{\rm H}/(6-5X_{\rm H}$) from Eq.~(\ref{DG_XH}) we get the fraction $f$ of the carbon atoms in aromatic rings
\begin{equation}
f = X_{sp^2} = \frac{1-4Q}{1+5Q} = \frac{1}{2} (2-3X_{\rm H}).
\end{equation}
Note that the above formalism implicitly takes into account the de-aromatisation of that part of the graphite lattice associated with the defect ({\it i.e.,} the conversion of carbon atoms from $sp^2$ to $sp^3$), and therefore $X_{sp^2}$ is a direct measure of the aromatic carbon atom fraction.

If we now consider the DG model in terms of stacked graphite sheets then we can see that defects (vacancies) in close proximity in adjacent sheets could lead to cross-linking between the layers. We note that nitrogen hetero-atoms could perhaps play an important role here because in nano-structured a-C:N films a strong cross-linking between graphitic planes is observed \cite[{\it e.g.},][]{2004PhilTransRSocLondA..362.2477F}. In any event, cross-linking leads to a reduction in the hydrogen atom content for a given value of $R$ by at least one hydrogen atom per defect. The loss of more than one hydrogen atom per defect, up to a maximum of three per defect, implies multiple cross-links per defect.  A minimum condition for all of the stacked DG layers to cross-link into a macroscopic structure is when each vacancy is associated with two cross-links, each forming an aliphatic bridge to one of the two adjacent sheets. We further assume that the cross-links on the $sp^3$ carbon atoms can, for steric hindrance reasons, only be on adjacent $sp^3$ carbon atoms. Thus, both CH$_2$ and CH groups can be present in the DG structure. If $L$ is the number of cross-links per defect ($0 \leq L \leq 3$) then the number of hydrogen atoms per defect is $(6-L)Q$  and we have that  
\begin{equation}
X_{\rm H}= \frac{(6-L)Q}{1+(5-L)Q}, 
\label{DG_XH_CL}
\end{equation}
\begin{equation}
X_{sp^3}= \frac{3Q}{1+(5-L)Q} = \frac{3 X_{\rm H}}{(6-L)}, 
\label{DG_Xsp3_CL}
\end{equation}
\begin{equation}
X_{sp^2}= \frac{1-4Q}{1+(5-L)Q} = \frac{(6-9X_{\rm H})-(1-X_{\rm H})L}{(6-L)}
\label{DG_Xsp2_CL}
\end{equation}
and 
\begin{equation}
R = \frac{3X_{\rm H}}{(6-L)-(9-L)X_{\rm H}} =  \frac{3X_{\rm H}}{(6 - 9X_{\rm H})-(1-X_{\rm H})L}, 
\label{DG_RvsXH}
\end{equation}
where we use the substituion  
\begin{equation}
Q = \frac{X_{\rm H}}{(6-L)-(5-L)X_{\rm H}} = \frac{X_{\rm H}}{(6 - 5X_{\rm H})-(1-X_{\rm H})L}
\end{equation}
We can also re-formulate Eqs.~(\ref{DG_coord}) and (\ref{DG_C_coord}) for the atomic coordinations for cross-linked DG structures using this substitution for $Q$ {\it {\it i.e.}},
\begin{equation}
{\bar m} = \frac{(6-3X_{\rm H})-(1-X_{\rm H})L}{2 -\frac{1}{3}(1-X_{\rm H})L}, 
\end{equation}
\begin{equation}
{\bar m_{\rm C}} = \frac{3[(6- 5X_{\rm H})-(1-X_{\rm H})L ]}{(6-L)(1-X_{\rm H})}, 
\end{equation}
which reduce to Eqs.~(\ref{DG_coord}) and (\ref{DG_C_coord}) when $L=0$.

Fig.~\ref{RvsXH_CH3} shows the results for the DG model of Eq.~(\ref{DG_RvsXH}), over its range of validity $0 \leq X_{\rm H} \leq 0.35$, assuming $L=2$, {\it {\it i.e.}}, that there are two cross-linking bonds per defect (carbon atom vacancy). As can be seen in Fig.~\ref{RvsXH_CH3}, this DG model gives reasonable agreement with the eRCN model and has the advantage that it can be used to describe hydrogen-poorer hydrocarbons extending towards a graphite-like structure. Also shown in Fig.~\ref{RvsXH_CH3} are the mean carbon atom coordination numbers, $3 \leq {\bar m}_{\rm C} \leq 4$, for the eRCN (solid red line) and DG (dashed red line) models.

\subsubsection{A compositional deconstruction of bulk DGs and their predicted infrared spectra}
\label{sect_DG_spectra}

In the same way as was done for eRCNs (see \S~\ref{structural_decomp}) we can deconstruct the DG network into its component structures. However, in this case the process is much simplified because the network consists of only aromatic $sp^2$ carbon atoms and aliphatic CH and CH$_2$ groups (at the vacancy sites). We then have, respectively, for the aromatic and aliphatic C-C bonds and the aliphatic CH and CH$_2$ groups concentrations:
$X_{\rm C\simeq C,DG} = \frac{3}{2} X_{sp^2}$,  
$X_{\rm C-C,DG} = 2 X_{sp^3}$, 
$X_{\rm CH_2,DG} = (3-L) \, X_{sp^3}$ and 
$X_{\rm CH,DG} = L \, X_{sp^3}$. 
Note that these equations indicate that the number of inter-atomic bonds is not equal to the total number of atoms in the structure. Thus, the atomic bond fractions have to be normalised by the total number of bonds, $X_{\rm bond}$, which is given by
\[
X_{\rm bond} =  \frac{3}{2} X_{sp^2}\ +\ 2\, X_{sp^3}\ +\ X_{\rm H}
\]
\begin{equation}
\ \ \ \ \ \ \ \ \ \, =  \frac{3}{2} X_{sp^2}\ +\ \left[2+\frac{(6-L)}{3} \right] X_{sp^3}
\label{eq_DG_bond_renorm}
\end{equation}
where we have substituted $X_{\rm H} = \frac{1}{3} (6-L) X_{sp^3}$. We then have for the DG model that the atomic bond fractions in the structure are
\begin{equation}
X_{\rm C\simeq C,DG} = \frac{3\, X_{sp^2}}{2\, X_{\rm bond}},  
\end{equation}
\begin{equation}
X_{\rm C-C,DG} = \frac{2 \, X_{sp^3}}{X_{\rm bond}}, 
\end{equation}
\begin{equation}
X_{\rm CH,DG} = \frac{L \, X_{sp^3}}{X_{\rm bond}} \ \ {\rm and}
\end{equation}
\begin{equation}
X_{\rm CH_2,DG} = \frac{(3-L) \, X_{sp^3}}{X_{\rm bond}}. 
\end{equation}

The only spectral bands that bulk DGs will show are therefore at $6.25$ and $6.67\,\mu$m due to aromatic C$\simeq$C  bonds, $3.45\,\mu$m due to aliphatic CH stretching, $3.42$ and $3.51\,\mu$m due to aliphatic CH$_2$ stretching modes, and at $6.9\,\mu$m due to aliphatic CH$_2$ bending ({\it e.g.,} see Table~\ref{spectral_bands}). However, for finite-sized particles the surface bonds will be terminated with H atoms and a large fraction of these will be aromatic CH bonds; as at the periphery of isolated polycyclic aromatic hydrocarbon (PAH) molecules. These DG particle, aromatic, surface CH bonds could provide  a catalytic r\^ole in molecular hydrogen formation in the ISM \cite[{\it e.g.},][]{2008ApJ...679..531R,2009ApJ...704..274L} and such processes, involving CH$_2$ sites, could lead to a significant fraction of surface (or PAH `edge') aliphatic CH$_2$ groups, in addition to the aromatic CH sites. The consequences of particle size for the DG model are discussed in detail in a following paper.

\section{The $3-13.5\,\mu$m spectra of hydrocarbon solids}
\label{sect_eRCN_DG_spectra}

\begin{figure*}
 \resizebox{\hsize}{!}{\includegraphics{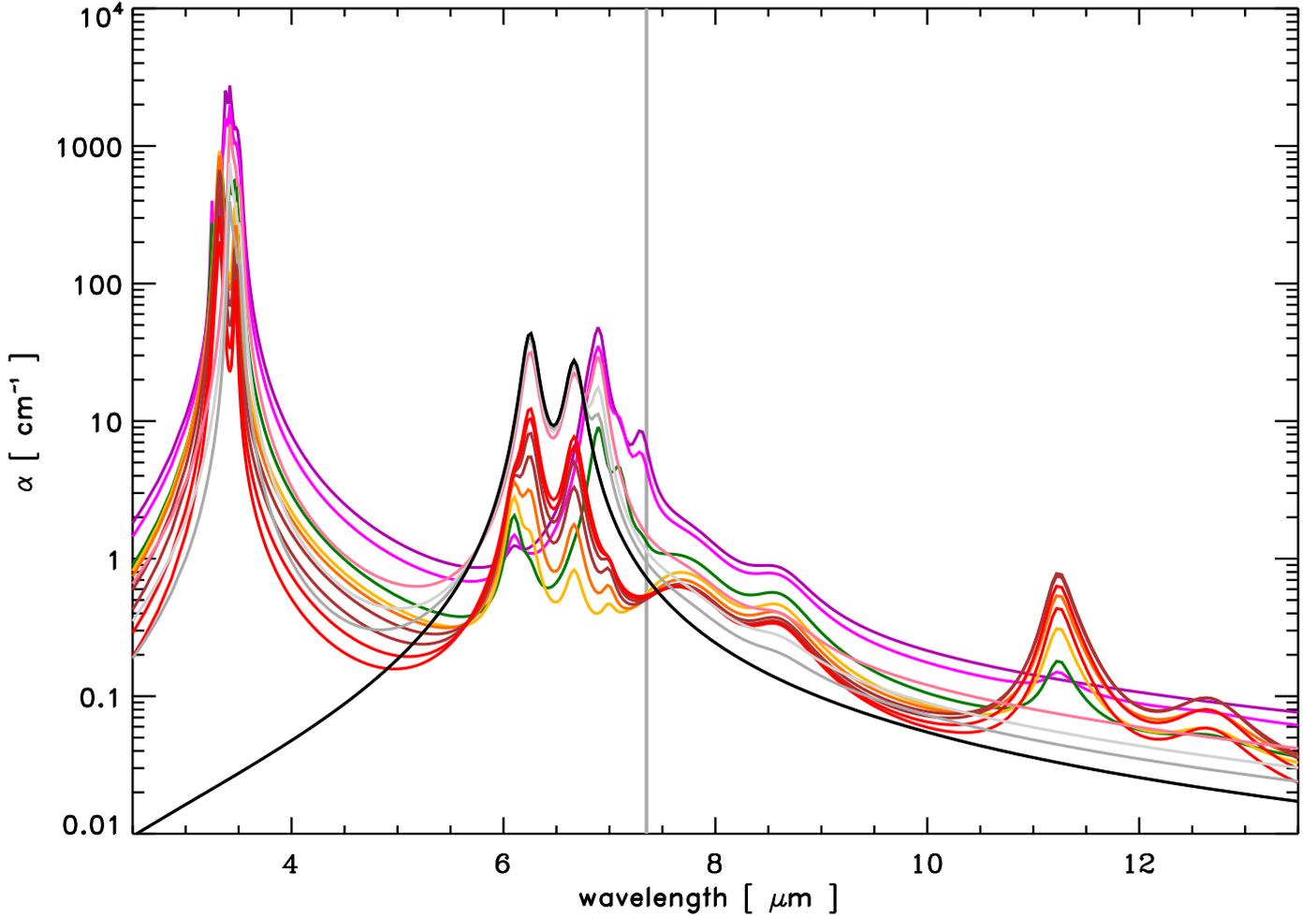}}
 \caption{The predicted  eRCN and DG spectra, presented as the absorption coefficient, $\alpha$, in the $2.5-13.5\, \mu$m region as a function of $E_{\rm g}$ calculated using the structural de-composition described in \S\,\ref{structural_decomp}, \S\,\ref{sect_eRCN_spectra} and \S\,\ref{sect_DG_model} and for the data in Table~\ref{spectral_bands}. The adopted colour-coding for this plot is from large band gap (2.67\,eV, purple) to low band gap (-0.1\,eV, black) with intermediate values in steps of 0.25\,eV from 2.5 (violet) to 0\,eV (grey) with the addition of the $E_{\rm g} = 0.1$ and $-0.1$\,eV cases (light grey and black, respectively), as per paper~II.  The bands with central positions long-ward of the vertical grey line ($\lambda(\nu_0) > 7.3\,\mu$m) are not yet well-determined by laboratory measurements.}
 \label{full_spectrum_eRCN_DG}
\end{figure*}

In Fig.~\ref{full_spectrum_eRCN_DG} we show the full, but provisional, spectra presented as the absorption coefficient, $\alpha$, for our eRCN/DG materials. These spectra are well-constrained for wavelengths less than $\simeq 7.3$\,$\mu$m where the band cross-sections have been determined experimentally. However, for the longer wavelength bands, and as indicated in the lower part of Table~\ref{spectral_bands} ({\it i.e.}, bands 20 to 27), we have introduced `place holder' bands with positions, widths and intensities, determined by comparison with a-C:H absorption coefficient data \citep[taken from][Fig.~14]{1986AdPhy..35..317R}, and also guided by the band positions and widths in the interstellar emission spectra \citep[{\it e.g.},][]{2001A&A...372..981V}, with the aim of allowing a qualitative exploration of the behaviour of the spectral variations of these materials in the ISM. However, some progress in this area has been made with the measurement of the absorption and emission spectra of some HAC materials and carbon nano-particles in the 3 to 20\,$\mu$m wavelength region \cite[{\it e.g.},][]{1997ApJ...490L.175S,2005ApJ...626..933D,2005ApJ...620L.135D,2006ApJ...653L.157H,2008ApJ...672L..81H,2008ApJ...677L.153H}.

We caution the reader in the use of the spectral bands longward of 7.3\,$\mu$m for the   detailed interpretation of astronomical spectra, at least until such time as these bands have been measured in the laboratory for a wide range of amorphous hydrocarbon solids.  Nevertheless, the adaptability of this model will allow for a rather easy revision of the complex index of refraction data once appropriate laboratory data become available.  

We note that the spectra shown in Fig.~\ref{full_spectrum_eRCN_DG} exhibit most of the IR bands observed in absorption and in emission in the ISM and that the model predicts a general weakening of the aliphatic bands, as $X_{\rm H}$ decreases, which is accompanied by a shift in character towards olefinic and aromatic-dominated bands. 
We also note that the relative intensities of the bands will depend upon particle size and not just on $X_{\rm H}$. This effect will be studied in detail in a follow-up paper.

\section{Astrophysical implications}
\label{sect_astro_implications}

The evolution of the spectral properties of carbonaceous dust, from its formation in the circumstellar shells around evolved stars to its transition into the ISM, has clearly been demonstrated \citep[{\it e.g.},][]{2003ApJ...589..419G,2007ApJ...662..389G,2007ApJ...664.1144S,2008A&A...490..665P}. The eRCN/DG  carbonaceous dust evolution model presented here, in which the nature of the dust is characterised by its hydrogen content, $X_{\rm H}$, and the carbon atom $sp^3/sp^2$ hybridisation ratio, $R = X_{sp^3}/X_{sp^2}$, allows a detailed exploration of the compositional transformations of these types of materials. 

We now examine some of the consequences of hydrogenated amorphous carbon dust evolution, as predicted by the eRCN and DG models, within the astrophysical context. However, a full examination will be presented in a follow-up paper, once the optical properties (paper~II) and the size dependence of these properties have been investigated.  

\subsection{Structural variations and spectral properties}
\label{sect_struct_spect_variations}

As a result of the effects of photo-darkening, aromatisation or `graphitisation' of a-C:H solids there is a transformation of carbon atoms from the $sp^3$ to the $sp^2$ hybridisation state that is necessarily accompanied by a net loss of H atoms from the structure \cite[{\it e.g.},][]{1996MNRAS.283..343D}. The eRCN/DG model predicts a maximum hydrogen content of about 62\%, which is in good agreement with experiment. Conversely, after extreme processing, it appears that practically all of the hydrogen can be removed from these materials. Ion irradiation experiments indicate a minimum H atom content of about 5\% but amorphous carbon materials formed in the laboratory can be made essentially hydrogen free and so it is expected that the lower limit is therefore `H-free' materials. 

In the ISM the large carbonaceous grains appear to have a CH$_2$/CH$_3$ ratio in the range $\sim 2-3$ \citep[{\it e.g.},][]{2002ApJS..138...75P,2005A&A...432..895D}. Fig.~\ref{fig_decomp_4} indicates that, for the eRCN model, this would be equivalent to a rather narrow hydrogen atom fraction range, {\it i.e.}, $\sim 53-55$\% ($\equiv X_{\rm H} = 0.53-0.55$). However, looking at Figs.~\ref{stretching_spectrum}, \ref{bending_spectrum} and \ref{full_spectrum_eRCN} we can see that it is the predicted spectra for a-C:H materials with $X_{\rm H} \geq 0.57$ that appear to be more consistent with the absorption  observed along the Galactic Centre \object{IRS6E}, \object{Cyg OB2 No. 12} and \object{IRS7} lines of site \citep[{\it e.g.},][]{2002ApJS..138...75P,2004A&A...423L..33D} and towards Seyfert 2 galaxy nuclei by \cite{2004A&A...423..549D} who identify the source material as being of a-C:H origin. We note that $X_{\rm H} \geq 0.57$ implies a CH$_2$/CH$_3$ ratio $\simeq 1$ (see Fig.~\ref{fig_decomp_4}), rather than the expected value of $\sim 2-3$. Thus, while the spectal agreement is rather good in the $3-4\,\mu$m region, it does not appear to be coherent with   the observed CH$_2$/CH$_3$ ratio. As indicated in Table~\ref{spectral_bands}, we needed to increase the intensity of {\em only} the $3.38\,\mu$m CH$_3$ band in order to obtain a good match to the observational data. Thus, the discrepancies noted here could be due to uncertainties in the band strengths and/or assignments and/or to particle size effects. 

The intrinsic CH, CH$_2$ and CH$_3$ groups responsible for the suite of absorption features around 3.4\,$\mu$m show a clear transformation from $sp^3$-dominated CH$_n$ bands to $sp^2$-dominated CH$_n$ bands as aromatisation proceeds and as $X_{\rm H} $ declines (see Figs.~\ref{fig_decomp_1} and \ref{fig_decomp_2}). The accompanying spectral variations shown in Figs.~\ref{full_spectrum_eRCN} and \ref{full_spectrum_DG} predict that the spectrum evolves towards one dominated by olefinic and aliphatic CH groups, with bands at 3.32 and 3.47\,$\mu$m in the case of eRCN, and only aliphatic CH and CH$_2$ bands at 3.42, 3.45, 3.47 and 3.51\,$\mu$m for the DG model. We note that the aromatic CH band at 3.28\,$\mu$m is not apparent in these spectra because of the lower abundance of the aromatic CH groups (with respect to olefinic and aliphatic CH groups) in the eRCN model and also because its lower intrinsic absorption strength (see Table~\ref{spectral_bands}). Aromatic CH bands are expected in the DG spectra but they are not included here, {\it i.e.}, the DG spectra do not yet take into account the likely hydrogenation of the aromatic surface nor the possibility of aromatic CH formation at defect sites. These aspects will be dealt with in detail in following paper.  

The evolution of the structural and spectral properties of a `single' carbonaceous material therefore appears to provide a good model for the evolution of amorphous hydrocarbons in interstellar and circumstellar media. With the addition of surface hydrogenation, and additional size effects, to the calculation of the spectra the DG model can be carried forward to a study of the most aromatic carbon particles in the ISM. These will necessarily be the smallest particles, UV-photo-darkened and stochastically-heated to high temperatures, as a result of which they will be dehydrogenated and aromatised. 

The $sp^3$ to $sp^2$ transition, as a result of UV photo-processing in the ISM, could possibly be counterbalanced by hydrogen atom addition to the structure as a result of (energetic) collisions \cite[{\it e.g.},][]{1996MNRAS.283..343D}. However, it is not yet clear that this `reverse' process of `aliphatisation' can occur because it requires that the incident H atoms are energetic enough to insert into aromatic rings and transform $sp^2$ C atoms into $sp^3$ sites. Given that the aromatic C$\simeq$C bond energy in benzene is of the order of 5\,eV, and that for an H atom to insert it must break at least one CC bond, the activation energy for this process is at least 5\,eV, equivalent to $> 5 \times 10^4$\,K, which would seem to indicate that there is a significant energy barrier to a-C/a-C:H hydrogenation in the ambient ISM. However, this is offset by the formation of an aromatic CH bond ($\sim 4.3$\,eV) and thus, if quantum mechanical tunnelling can occur, the barrier to aliphatisation could be as low as 0.7\,eV ($\approx 8000 $\,K). These types of interactions have been well-studied experimentally and theoretically by \cite{2000ApJ...528..841D} and \cite{2008ApJ...682L.101M,2010ApJ...718..867M}, within the context of carbonaceous dust evolution in dense regions, 
and we therefore do not enter into any further speculation on the aliphatisation issue here. 

\subsection{H$_2$ formation in PDRs via a-C:H de-hydrogenation?}

Recent observations and their interpretation appear to require rather efficient molecular hydrogen formation at `elevated' temperatures in photon-dominated regions \citep[PDRs, {\it e.g.},][]{2004A&A...414..531H,2011A&A...527A.122H}.  \cite{2008ApJ...679..531R} and \cite{2009ApJ...704..274L} propose that H$_2$ formation can be catalysed by reactions involving the chemical trapping of H atoms at the periphery of PAH molecules in the ISM. In a rather similar vein, we note that in a-C:H materials `UV-photolysis' or ion irradiation, resulting in an aromatisation, induces H atom loss from the structure and is associated with the {\it in situ} formation of molecular hydrogen, as studied and modelled by \cite{1984JAP....55..764S}, \cite{1989JAP....66.3248A}, \cite{1996MCP...46...198M} and \cite{2011A&A...529A.146G}.

Initially, at least, the formation of H$_2$ and CH bonds occurs as interstitial H atoms, with binding energies of the order of $0.05-0.2$\,eV ($\equiv 600-2400$\,K), become mobile, react and are lost from the solid, consistent with an observed increase in the aliphatic 3.4\,$\mu$m band intensity for temperatures above 573\,K, a process that rapidly terminates at higher temperatures \cite[{\it e.g.},][]{1984JAP....55..764S,1989ApPhL..54.1412S,1996MNRAS.283..343D}. 

Thus, the formation of H$_2$ associated with the aliphatic to aromatic transformation of amorphous hydrocarbon grains should occur in PDRs where the abundant UV photons will progressively de-hydrogenate any H-rich a-C:H material that  has been `ablated'  or `advected' from the nearby, denser molecular regions. In this case it is the de-hydrogenation of the smallest a-C(:H) particles that could contribute most to the formation of H$_2$ because of their relatively large abundance and the fact that they will also undergo stochastic heating to temperatures of the order of hundreds of degrees K. However, their stock of hydrogen is rather low. The reverse process of H atom addition to the aromatic component of these small carbon grains  \citep[{\it e.g.},][]{2008ApJ...679..531R,2009ApJ...704..274L} could offset H atom loss by UV photo-dissociation. Thus, an equilibrium between the UV and/or thermal processing of small a-C:H particles in PDRs and H atom addition could be a viable source of H$_2$ formation at `high' dust temperatures where more classical models for H$_2$ formation appear to be inadequate (see paper~II for a quantification of this process).

\subsection{a-C:H decomposition products}

It is interesting to note that the aromatisation of a-C:H materials will yield decomposition products as the materials evolve. The principal product, as discussed above, will be molecular hydrogen. However, it is likely that other species will also be released as the structure re-adjusts to a new composition. As was clearly shown by \cite{1984JAP....55..764S}, a-C(:H) loses its hydrogen not just as H$_2$ but also as hydrocarbon molecules, which leads to significant mass loss from the samples during thermal annealing. The release of `molecular' daughter species is likely to become more pronounced as the particle size decreases as suggested by \cite{1996MNRAS.283..343D}, \cite{1997ApJ...490L.175S}, \cite{2000ApJ...528..841D} and \cite{2009ASPC..414..473J}.  

\cite{2005A&A...435..885P} proposed that the small hydrocarbons, such as CCH, c-C$_3$H$_2$, C$_4$H that they observed in PDRs derive from precursor PAHs. This hypothesis was proposed by \cite{1996ApJ...472L.123S} and extended by \cite{2009ASPC..414..473J} who both postulated that the precursors for these `PAHs' themselves derive from the photo-thermal de-construction of small a-C(:H) particles \citep[as is indeed supported by the laboratory work of ][]{1984JAP....55..764S} in the transition from dense, molecular regions of the ISM to PDRs. These precursor `PAHs' may rather resemble the locally aromatic polycyclic hydrocarbons (LAPH), cousins of PAHs, that were proposed by \cite{2003ApJ...594..869P} as a carrier of the IR emission bands. 

Here we follow the suggestion of \cite{1996ApJ...472L.123S} and \cite{2009ASPC..414..473J} that a-C(:H) materials in the photon-dominated ISM follow a continuous evolutionary sequence that schematically can be represented as:  \\[0.2cm] a-C:H $\rightarrow$ a-C $\rightarrow$ `PAHs'/`LAPHs' $\rightarrow$ small hydrocarbons. \\[0.2cm] 
The exact evolutionary track of the carbonaceous particles will depend upon their size. This is clearly only a part of what must be a complete carbon cycle and so the above needs to be complemented by `return journey', {\it i.e.}, a re-formation of a-C(:H) materials somewhere in the ISM, perhaps this can occur by a-C(:H) accretion onto pre-existing grains as proposed long ago by \cite{1990QJRAS..31..567J}, more recently by \cite{2009ASPC..414..473J} and alluded to by \cite{2011A&A...530A..44J}.

\section{Predictions of the eRCN/DG model}
\label{sect_predictions}

Here we summarise some of the more important predictions of the eRCN/DG model: 
\begin{enumerate}
  \item The determination of a-C:H/a-C structures, containing only $sp^2$ and $sp^3$ C atoms and H atoms, and their related spectra based on only the critical H atom content, $X_{\rm H}$ (or equivalently the band gap energy $E_{\rm g} $(eV) $\simeq 4.3 X_{\rm H}$), as a result of UV photon-driven and/or thermal annealing, where $sp^3$ C atoms are transformed into $sp^2$ C atoms as annealing proceeds. 
  \item With annealing the a-C:H spectra `evolve' more rapidly for the hydrogen-rich materials, {\it i.e.}, $0.5 < X_{\rm H} < 0.62$, than for the hydrogen-poorer materials, due to the loss of hydrogen from the structure and the necessary major structural re-arrangement that is required to compensate for this loss. 
  \item The expected spectrum of the large carbonaceous grains in the ISM is given by the purple spectra in  Figs.~\ref{stretching_spectrum} to \ref{full_spectrum_eRCN} and \ref{full_spectrum_eRCN_DG}; 
  based on the fit to the observed interstellar absorption in the $3.2-3.6\,\mu$m region the spectra indicate  $X_{\rm H} \gtrsim 0.57$ with a CH$_2$/CH$_3$ ratio $\approx 0.5-1.3$ (Fig.~\ref{fig_decomp_4}).
  \item That the photo-thermal dehydrogenation/decomposition of a-C:H provides an efficient route to molecular hydrogen formation, and to the production of small daughter hydrocarbons ({\it e.g.}, CCH, c-C$_3$H$_2$, C$_4$H, {\it etc}.), in regions where their formation would otherwise appear to be inhibited. 
\end{enumerate}

\section{Limitations of the eRCN/DG model}
\label{sect_limitations}

The eRCN and DG structural models for a-C(:H) materials, and the spectral predictions that they engender, appear to work rather well. However, it must be remembered that they are only an approximation of reality and so they do have their limitations, which principally seem to be that: 
\begin{itemize}
  \item The spectral bands long-wards of $7.3\,\mu$m are seemingly not yet well-enough determined experimentally for a-C(:H) materials. Caution should therefore be exercised in their use to interpret astrophysical observations. 
  \item The eRCN/DG model cannot currently deal with `dangling bonds' and therefore with dehydrogenated a-C(:H) materials that might include a significant fraction of them. Given that `dangling bonds' resulting from dehydrogenation will generally lead to a structural re-arrangement (as the eRCN/DG model indeed predicts) this is probably of little concern for the bulk composition.  However, as \cite{2011ApJ...737L..44D} show, their effects could be rather important. 
  \item The model does not consider the role and consequences of interstitial H atoms \citep[{\it e.g.},][]{1989ApPhL..54.1412S,1996MNRAS.283..343D,2011ApJ...737L..44D} because they do not bond into the network. 
  \item $sp^2$ clustering into chain or three-dimensional cage-like structures is currently not considered but could be particularly important for small particles where the inclusion of C$_5$ pentagons into the aromatic clusters would quite naturally lead to shell-like, ``fullerene-type'' clusters. 
  \item $\equiv$C--H and --C$\equiv$C-- $sp^1$ sites, and chains thereof, are not included in the network calculations. 
  To first order, they are expected to be of low abundance ($\lesssim 5$\%) and this may not be a major limitation. 
  \item This version of the eRCN/DG model does not take into account the presence of hetero-atoms such as O and N but they could, in principle, be included in a modified version of the eRCN model.
\end{itemize}

\section{Conclusions}
\label{sect_conclusions}

In this paper, the first in a series, we present a self-consistent model for the evolution of the structural and spectral properties of hydrogenated amorphous carbon solids, a-C(:H), which is based on the Random Covalent Network (RCN) and defective graphite (DG) models.
We show that the evolution of a `single' hydrogenated amorphous carbon material appears to provide an excellent model for the evolution of these types of carbonaceous dust materials in interstellar and circumstellar media. 
In particular, this model is consistent with the following laboratory and astronomical observations:
\begin{itemize}
  \item The expected CC and CH$_n$ bonding structures and chemical compositions of the suite of hydrogenated amorphous carbons (a-C:H/a-C).  
  \item The compositional and spectral evolutionary transition from aliphatic-dominated materials to olefinic-dominated (and eventually to aromatic-dominated) with increasing exposure to UV photons and with decreasing grain size.
  \item The formation of molecular hydrogen, as a result of `annealing' after photon- or incident ion-induced dehydrogenation, at dust temperatures higher than those for `classical' surface formation processes. 
  \item The photo-thermal driven `aromatisation' production of daughter, hydrocarbon molecular species including: locally aromatic polycyclic hydrocarbons (LAPHs), `PAHs' and molecular fragments such as CCH, c-C$_3$H$_2$, C$_4$H. 
\end{itemize}

Note that the model here does not consider the effects of particle size on the compositional and spectral properties. However, the modelling approach presented here is clearly good for the analysis of bulk materials, and could be valid for particles as small as 10\,nm, where their bulk properties dominate over those of their surfaces. A detailed presentation of particle size effects will be presented in forthcoming paper.  

In the following paper (paper~II) we derive the optical properties of a-C(:H) materials from EUV energies to cm wavelengths. 


\begin{acknowledgements} 
I would like to thank the referee, Walt Duley, for many valuable suggestions. This research was, in part, made possible through the financial support of the Agence National de la Recherche (ANR) through the program {\it Cold Dust} (ANR-07-BLAN-0364-01). 
\end{acknowledgements}


\bibliographystyle{aa} 
\bibliography{biblio_HAC.bib} 




\appendix

\section{Aromatic cluster characterisation}
\label{aromatics}

In the study of polycyclic aromatic hydrocarbon (PAH) molecules and polycyclic aromatic clusters in hydrogenated amorphous carbon solids (a-C:H or HAC) we need to characterise the constituent aromatic structures in terms of their  number of carbon atoms, $n_{\rm C}$, number of aromatic, six-fold rings, $N_{\rm R}$, or coordination number, $m$. 
It has been shown that $N_{\rm R}$ is a function of the band gap, $E_{\rm g}$, of a given hydrogenated amorphous carbon, {\it i.e.,} $N_{\rm R} = [5.8/E_{\rm g}({\rm eV})]^2$, and that $E_{\rm g}$ can be expressed as a function of $X_{\rm H}$, {\it i.e.} $E_{\rm g} \sim 4.3 X_{\rm H}$, the aromatic coherence length, $L_{\rm a}({\rm nm}) = [0.77/E_g({\rm eV})]$, a measure of the aromatic domain size, also depends on the band gap \cite[{\it e.g.},][]{1986AdPhy..35..317R,1991PSSC..21..199R}. 

In section \ref{section_aromatics} we introduced an `aromatic cluster parameter', $Z$, which is a function of $N_{\rm R}$ and is simply the number of constraints, $N_{\rm con}$, per carbon atom for the given cluster, {\it i.e.},
\begin{equation}
Z = \frac{N_{{\rm con},N_{\rm R}}} {n_{\rm C}}
\end{equation}
where $n_{\rm C}$ is the number of carbon atoms per aromatic cluster. 
For rather small and compact aromatic clusters such as pyrene, perylene, {\it etc.}  \cite{1990MNRAS.247..305J} showed that $Z = (\frac{5}{2}N_{\rm R}+12)/(3N_{\rm R}+4)$ and for linear aromatic systems such as naphthalene and anthracene that $Z = (5N_{\rm R}+7)/(4N_{\rm R}+2)$. For the linear clusters the expressions are valid for all $N_{\rm R}$. However, for compact clusters the expressions give a reasonable fit for $N_{\rm R} \leq 10$ which is applicable to most of the aromatic clusters that occur in amorphous hydrocarbons \citep{1986AdPhy..35..317R} and therefore, by inference, in our eRCNs. Additionally, as pointed out by \citep{1986AdPhy..35..317R}, the compact clusters will be more stable that the row clusters. 

The relevant expressions for linear and compact aromatic clusters are shown in Table~\ref{PAH_cluster_params}. 
The compact cluster expression given in the upper part of the table is in fact that for bi-linear clusters, {\it i.e.}, two parallel and connected linear structures. The middle part of the table gives the expressions for clusters consisting of a given number of rows of linear structures, $N_{\rm row}$, where in this case the values of $N_{\rm R}$ are restricted to $N_{\rm R} \geq p N_{\rm row}$ where $p$ is an integer $\geq 1$. The expressions for compact clusters that give reasonable fits for all $N_{\rm R}$ are given in the lower part of the table.

\begin{table}
{\footnotesize 
\caption{The expressions for (polycyclic) aromatic cluster structures.}
\begin{center}
\begin{tabular}{lll}
                       &                             &                                       \\[-0.35cm]
\hline
\hline
                       &                             &                                       \\[-0.35cm]
  {Term}           &  {Linear clusters}      &  {Compact clusters}               \\[0.05cm]
\hline
                       &                             &                                       \\[-0.35cm]
  No. of C atoms, $n_{\rm C}$  & $4N_{\rm R}+2$              &  $3N_{\rm R}+4$                       \\[0.05cm]
  Coordination no.     & $2N_{\rm R}+4$              &  $N_{\rm R}+6$                        \\[0.05cm]
                       & \ \ $= \frac{1}{2}n_{\rm C}+3$      &  \ \ $= \frac{1}{3}(n_{\rm C}+14)$            \\[0.05cm]
  No. of rings         & $\frac{1}{4}(n_{\rm C}-2)$          &  $\frac{1}{3}(n_{\rm C}-4)$                   \\[0.05cm]
  No. of constraints   & $5N_{\rm R}+7$              &  $\frac{5}{2}N_{\rm R}+12$            \\[0.05cm]
                       & \ \ $= \frac{1}{6}(5n_{\rm C}+52)$  &  \ \ $= \frac{1}{4}(5n_{\rm C}+18)$           \\[0.05cm]
\hline
                       &                             &                                       \\[-0.35cm]
\multicolumn{3}{l}{Analytical fit for $N_{\rm row}$ linear clusters}                     \\[0.05cm]
  No. of C atoms, $n$  &                             &  $2[N_{\rm R}+\frac{N_{\rm R}}{N_{\rm row}}+N_{\rm row}]$ \\[0.05cm]
  Coordination no.     &                             &  $2[\frac{N_{\rm R}}{N_{\rm row}}+N_{\rm row}+1]$  \\[0.05cm]
  No. of rings         &                             &  $N_{\rm row}(n-N_{\rm row})/[2(N_{\rm row}+1)]$   \\[0.05cm]
  No. of constraints   &                             &  $5(\frac{N_{\rm R}}{N_{\rm row}}+N_{\rm row}+1)$  \\[0.05cm]
\hline
 \multicolumn{3}{l}{Analytical fit for all compact clusters}                             \\[0.05cm]
  No. of C atoms, $n$  &                             &  $2N_{\rm R}+3.5\surd N_{\rm R}+0.5$  \\[0.05cm]
  Coordination no.     &                             &  $3.5 \surd N_{\rm R}+2.5$            \\[0.05cm]
                       &                             &  \ \ $= \frac{7}{8} [(\frac{7}{8})^2+\frac{1}{2}(n-\frac{1}{2})]^\frac{1}{2} - \frac{9}{16}$                     \\[0.05cm]
  No. of rings         &                             &  $\{[(\frac{7}{8})^2+\frac{1}{2}(n-\frac{1}{2})]^\frac{1}{2} - \frac{7}{8} \}^2$   \\[0.05cm]
  No. of constraints   &                             &  $\frac{5}{4}(7 \surd N_{\rm R}+5)-3$ \\[0.05cm]
\hline
\hline
                       &                             &                                       \\[-0.35cm]
\end{tabular}
\end{center}
\label{PAH_cluster_params}
}
\end{table}

\begin{figure}
 \resizebox{\hsize}{!}{\includegraphics{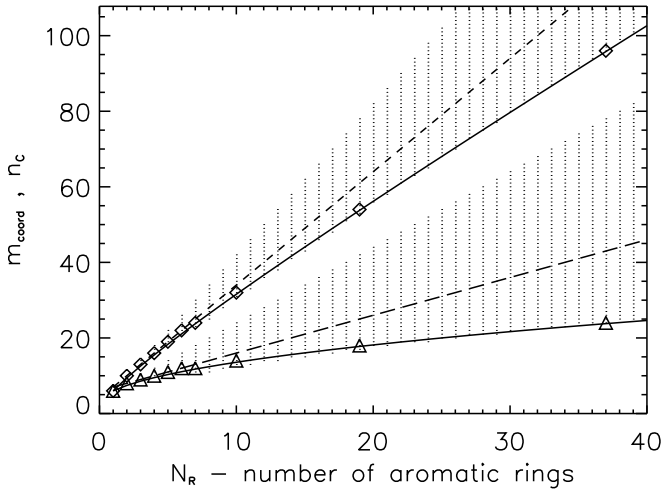}}
 \caption{Number of carbon atoms $n_{\rm C}$ (upper curves, diamonds) and coordination number $m_{\rm coord}$ (lower curves, triangles) for the most compact aromatic clusters possible (solid lines). The dashed lines show the simple fits to the data for general aromatic clusters.}
\label{fig_m_nc}
\end{figure}

The expressions for the number of carbon atoms per aromatic cluster, $n_{\rm C}$, and the coordination number $m_{\rm coord}$ of the cluster are plotted in Fig.~\ref{fig_m_nc}. The dashed lines in this figure show the simple fits to the data for general aromatic clusters, which are valid for the most compact clusters containing up to about 10-rings but deviate for larger systems (the expressions for compact clusters are taken from the upper portion of Table~\ref{PAH_cluster_params}). The areas indicated by the vertical rows of dots show the values for {\em all possible} aromatic clusters with a given number of rings, {\it i.e.}, the values for all possible isomers. The upper limits to the dots are for the most extended linear acenes (naphthalene, anthracene, tetracene, pentacene, etc.). Note that the simple expressions, represented by the dashed lines, give a very good representation of moderately compact clusters over the entire range of $N_{\rm R}$ considered here.

\begin{figure}
 \resizebox{\hsize}{!}{\includegraphics{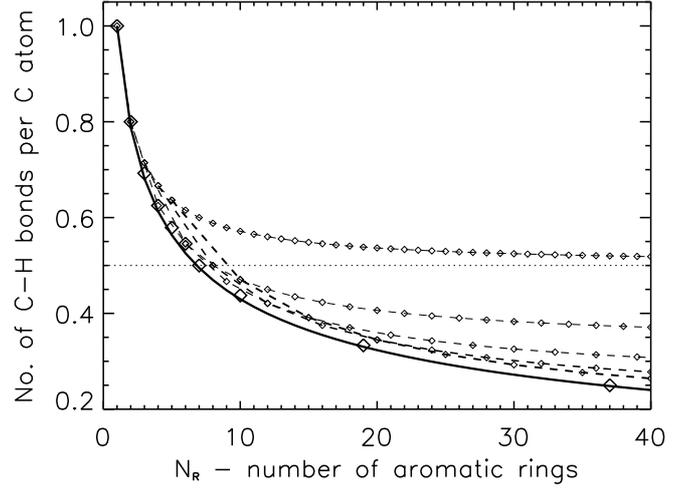}}
\caption{Ratio of the coordination number $m$ to the number of carbon atoms $n_{\rm C}$ in aromatic clusters, {\it i.e.}, the maximum possible number of peripheral aromatic cluster C--H bonds per carbon atom. See the text for a full description of the plotted data.}
\label{fig_m_per_c_atom}
\end{figure}

The coordination number per carbon atom in the aromatic cluster is plotted in Fig.~\ref{fig_m_per_c_atom}.
In this figure the large diamonds show the values for the most compact aromatic clusters possible, {\it i.e.}, for the large clusters these are the equivalent to the coronene, ovalene, circumcorenene, etc. structures. The small diamonds and their connecting dashed lines show the values for the multilinear acenes with $N_{\rm row} = $ 1, 2, 3, 4 and 5 (from top to bottom). The small diamonds mark the actual cluster values and the dashed lines are calculated using the multilinear acene expressions (given in the middle of Table~\ref{PAH_cluster_params}). The horizontal dotted line indicates the limiting values (0.5) for simple linear acenes ($N_{\rm row} = 1$). Note that the compact cluster expression in the upper portion of Table~\ref{PAH_cluster_params} is equivalent to the multilinear acene expression with $N_{\rm row} = 2$. The thick solid line gives a fit for the most compact aromatic clusters calculated using the expressions in the lower portion of Table~\ref{PAH_cluster_params}.

\section{eRCN model extension possibilities}
\label{extensions}

The mathematical basis of the RCN and eRCN models  \citep[{\it e.g.},][]{1979PhRvL..42.1151P,1980JNS...42...87D,1983JNCS...57..355T,1988JVST....6.1778A,1990MNRAS.247..305J} is a pair of equations with three inter-dependent variables, {\rm viz.}, 
\begin{equation}
X_{\rm H} + X_{sp^2} + X_{sp^3} = 1 
\end{equation}
and
\begin{equation}
aX_{\rm H} + bX_{sp^2} + cX_{sp^3} = 3. 
\end{equation}
Solutions to these equations are found by combining or `collapsing' two of the variables into a single one, {\it i.e.}, $X_{sp^3}/X_{sp^2} \equiv R$, and solving for $R$ as a function of $X_{\rm H}$. 
It is therefore evident that the addition of any new structural component into the system introduces a new variable, which renders the resulting equations unsolvable because we then have two equations and four variables. 
This limitation could be circumvented in several ways:
\begin{enumerate}
  \item By a collapsing of three of the variables into one, {\it i.e.}, $X_{sp}/(X_{sp^3}+X_{sp^2})= P$. However, this leads to ambiguities because of the degeneracy between the variables in the denominator and therefore to no useful or physically-realistic solutions. 
  \item By fixing the concentration of the new component. This is a useful strategy but requires that the atomic fraction of the new component, $W$, is well-constrained by measurements and also that the number of involved atoms is invariant and independent of the network structure, {\it i.e.}, no atomic loss, or transformation, of component $W$ is allowed. New carbon atom hybridisation or polyatomic structures, such as $sp^1$ C atoms or five-fold C$_5$ aromatic rings, do not fulfil this condition and therefore cannot be treated in this way. 
  
  The addition of hetero-atoms, such as O and N, should be possible using this approach, provided that their atomic abundances remain fixed. However, fixing the abundance of O and/or N atoms, and then treating their contribution as a constant, does not correctly take into account their accommodation into the structure and does not lead to realistic or useful solutions. 
  \item By the elimination of one of the variables. For example, in order to introduce $sp^1$ C atoms it is possible to eliminate either the $X_{sp^3}$ or the $X_{sp^2}$ component and then solve as a function of $X_{\rm H}$ by combining the remaining two variables. However, in neither of these particular cases does this result in physically-meaningful solutions, which is not surprising since this treatment imposes unrealistic network constraints, {\it i.e.}, networks consisting of only $sp^1$ and $sp^2$ atoms, with 2- and 3-fold coordination, or of only $sp^1$ and $sp^3$ atoms, with 2- and 4-fold coordination. The readers are left to verify this limitation for themselves. 
\end{enumerate}
In conclusion, it does not yet appear possible to add another carbon atom or hetero-atom component to the eRCN model in a physically-realistic way.


\end{document}